# Recent Results of the CMS Experiment

**Tommaso Dorigo**[1]

**on behalf of the CMS Collaboration**

*INFN, Sezione di Padova*
*Via F.Marzolo 8, 35131 Padova, Italy*
*E-mail:* `tommaso.dorigo@gmail.com`

The CMS collaboration has recently produced results of a number of searches for new physics processes using data collected during the 2011 run of the Large Hadron Collider. Up to 5 inverse femtobarns of proton-proton collisions at 7 TeV centre-of-mass energy have been used to search for the standard model Higgs boson in five different decay modes, divided in 42 independent sub-channels. The combination of the results has allowed CMS to set 95% confidence-level limits on the Higgs boson mass, constraining it to lay in the region *114.4<$M_H$<127 GeV* or *$M_H$>600 GeV*. An excess of events with a local significance of 3.1σ is observed for *$m_H$=124 GeV*[2]; the global significance of observing such an effect anywhere in the search range 110-600 GeV is estimated to be 1.5σ. A number of signatures of supersymmetric particles have also been investigated, significantly restricting the parameter space of natural low-scale theories. A search for the rare decays of neutral bottom mesons to muon pairs, $B_s \to \mu\mu$ and $B_d \to \mu\mu$, has achieved the tightest limits to date[3], and is approaching the sensitivity to measure the standard model branching ratios. As it happens, though, the highly informative results extracted from 2011 data produce more questions than answers; this doubles expectations for the 2012 run of LHC, which will conclusively answer several of them.

---

[1] Speaker
[2] The quoted results on combined Higgs boson searches have been updated by a recent publication[1].
[3] The LHCb collaboration has since produced a slightly tighter limit which is now the world's best on this search[2].

## 1. Introduction

2012 is arguably a crucial year for the Large Hadron Collider (LHC) experiments in particular, and for high-energy physics in general. After the start-up of the LHC at the CERN laboratories in Geneva on September 10$^{th}$ 2008, and the subsequent September 19$^{th}$ incident which caused a full additional year of delay in the schedule of the machine[4], physicists working at the LHC experiments have been busy with the exploitation of data coming first from low-luminosity, low-energy collisions, and then gradually higher energy and beam intensity (see Fig. 1). While insufficient to produce significant advancements in our knowledge of fundamental physics, the very early, $\sqrt{s}$ = 900 GeV data delivered by the LHC in 2009 allowed important calibrations of the detectors and data acquisition chain through the re-discovery of known standard model (SM) signals. In contrast, the 40 inverse picobarns of 7-TeV collisions delivered in the fall of 2010 were successfully used by the CMS collaboration to tighten the existing bounds on supersymmetric (SUSY) parameters[3-7], as well as to discover new features of low-energy quantum-chromo-dynamics (QCD) production[8], and to add the 7-TeV measurement point to several cross section graphs[9-12]. However, it was the data collected in 2011 which brought a significantly larger amount of information on many new physics models. And 2012 promises to allow a further qualitative step forward.

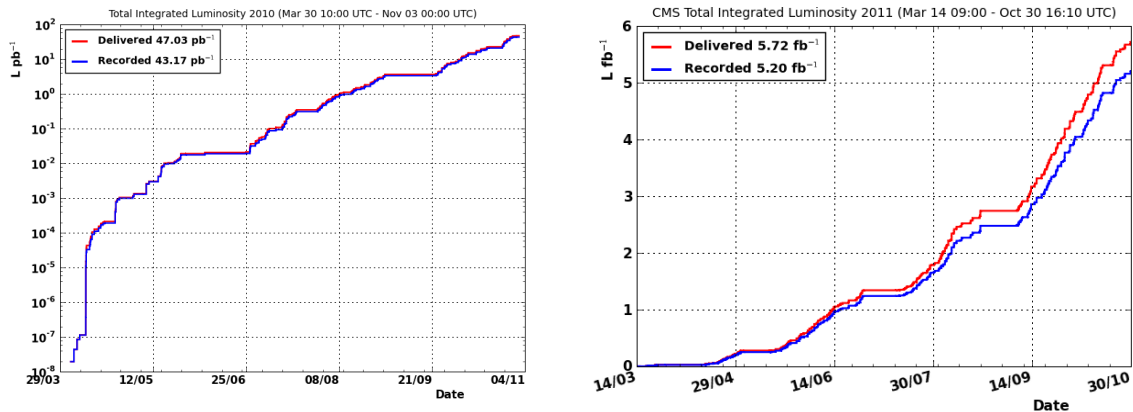

Figure 1. *Left: Integrated luminosity (in inverse picobarns, on a semi-logarithmic scale) delivered by the LHC (red) and recorded by the CMS experiment (blue) during the 2010 7-TeV proton-proton run, as a function of date of the year. Right: Integrated luminosity (in inverse femtobarns) delivered by the LHC (red) and recorded by the CMS experiment (blue) during the 2011 7-TeV proton-proton run, as a function of date of the year.*

In this paper we offer a review of some of the most interesting searches performed with 2011 data by the CMS experiment. As the reader will see, many of the results configure themselves as open questions: is the $B_s$ meson decaying to muon pairs at the rate predicted by the standard model, or at a higher rate compatible with the exchange of virtual massive SUSY mediators ? Does the Higgs boson exist ? Is the Higgs boson mass compatible with the indications of the existing global fits to electroweak observables? Can gluinos with mass below 1 TeV be excluded for most of the SUSY parameter space, throwing a monkey wrench into the inner workings of the most studied extensions of the standard model ?

---

[4] See *e.g.* http://press.web.cern.ch/press/PressReleases/Releases2008/PR14.08E.html

Most of the above questions are very likely to certain to receive a conclusive answer this year, when the LHC is scheduled to run at a centre-of-mass energy of 8 TeV, and is foreseen to deliver over 15 inverse femtobarns of proton-proton collisions to the ATLAS and CMS detectors. The 1-TeV increase in total collision energy with respect to 2011 may not sound like a large improvement; yet several of the most interesting final states which might produce discoveries of new physics, such as *e.g.* a new heavy Z' gauge boson, correspondingly increase their cross sections quite significantly because of their dominant production via quark-antiquark initial states; even precision physics measurements in the top quark sector will benefit from the over 50% enhancement of top-pair production rate, effectively increasing the overall number of analyzable top quarks by a factor of 5 with respect to 2011.

The structure of this document is as follows. We provide some detail of the CMS experiment and of the 2011 data taking campaign in section 2. Section 3 describes some of the most significant measurements produced by CMS in electroweak and top quark physics. We offer a discussion of selected B physics results in section 4. In section 5 the Higgs boson searches in the full 2011 dataset are described in detail. Section 6 is devoted to a brief summary of CMS searches for SUSY processes. Section 7 discusses two recent results for exotic new physics processes. We provide our conclusions in section 8.

## 2. The CMS Detector and the 2011 Run

CMS –an acronym for *Compact Muon Solenoid*– is a multi-purpose magnetic detector designed to study proton-proton collisions delivered by the CERN Large Hadron Collider. The detector is located in a underground cavern at a depth of 100m at the site of Cessy, near the border of France and Switzerland. Particles emitted in hard collisions at the center of CMS cross in succession a silicon tracker, electromagnetic and hadron calorimeters, a solenoid magnet, and muon drift chambers embedded in the solenoid iron return yoke. A drawing of the CMS detector is shown in Fig. 2.

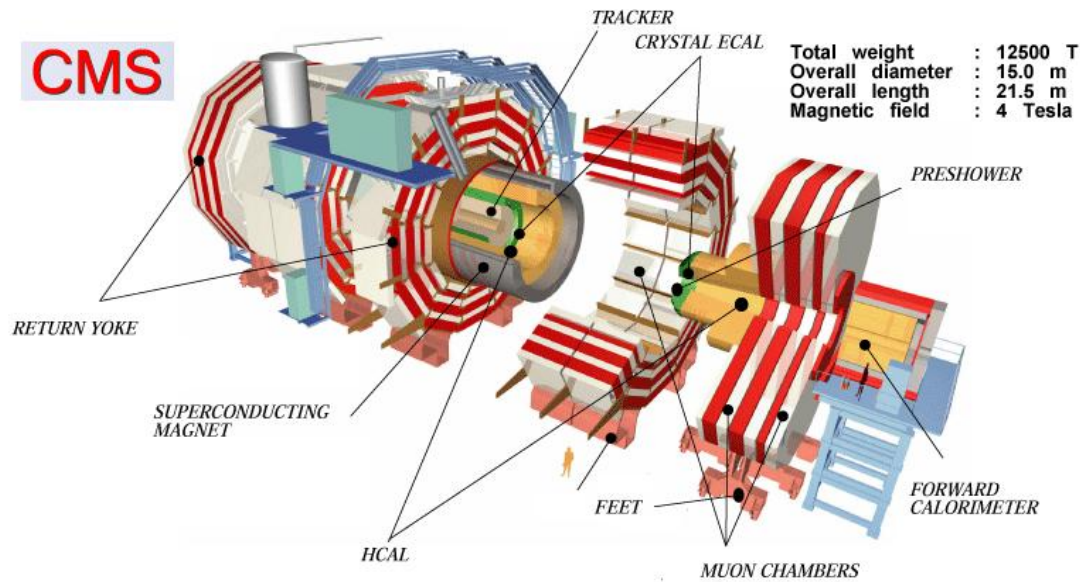

Figure 2. *An exploded view of the CMS detector, showing the outer muon chambers (white) embedded in iron (red). Internally can be seen the calorimeter system (ECAL, in green, and HCAL, in orange). The tracker is located in the core of the central barrel.*

## 2.1 Overview of the CMS Detector

The momenta of charged particles emitted in the collisions at the center of the CMS detector are measured using a 13-layer silicon pixel and strip tracker; 66 million silicon pixels of dimensions 100x150 μm are arranged in three barrel layers, and are surrounded by 9.6 million 180 μm-wide silicon strips arranged in additional concentric barrels in the central region and disks in the endcap region. In order to allow a precise measurement of charged particle momenta, the silicon tracker is immersed in the 3.8 T axial field produced by a superconducting solenoid. The tracker covers the pseudorapidity range $|\eta| < 2.5$, where pseudorapidity is defined as $\eta = -\ln \tan \theta/2$ and $\theta$ is the polar angle of the trajectory of a particle with respect to the direction of the counter-clockwise proton beam.

    Surrounding the tracker are an electromagnetic calorimeter (ECAL) composed of lead tungstate crystals, and a brass-scintillator hadron calorimeter (HCAL). These detectors are used to measure the energy of incident particles from the produced electromagnetic and hadronic cascades; they consist of a barrel assembly covering the central region, plus two endcaps covering the solid angle for particles emitted at lower angle with respect to the beams direction. The ECAL and HCAL extend to a pseudorapidity range of $|\eta| < 3.0$; at still smaller angles particles emitted in the collision encounter a steel/quartz-fiber Cherenkov forward detector (HF) which extends the calorimetric coverage to $|\eta| < 5.0$.

    The outermost component of the CMS detector is the muon system, consisting of four layers of gas detectors placed within the steel return yoke. The CMS muon system

performs a high-purity identification of muon candidates and a stand-alone measurement of their momentum, and in combination with the inner tracker information provides a high-resolution determination of muon kinematics. More detail on the CMS detector is provided elsewhere[13].

CMS collects data with a two-level trigger system. Level 1 is a hardware trigger based on custom-made electronic processors that receive as input a coarse readout of the calorimeters and muon detectors and perform a preliminary selection of the most interesting events for data analysis, with an output rate of about 100 kHz. Level 2, also called "High-Level Trigger" (HLT), uses fine-grained information from all sub-detectors in the regions of interest identified by Level 1 to produce a final decision, selecting events at a rate of about 300 Hz by means of speed-optimized software algorithms running on commercial computers.

## 2.2 The 2011 Run of the LHC

The 2011 proton-proton run of the LHC started on March 14$^{th}$ and terminated on October 30$^{th}$. In the course of the seven months of data-taking CMS acquired a total of 5.3 inverse femtobarns of integrated luminosity; 5.0 of these were collected with all the CMS subdetectors fully operational. Most of the results discussed in this document have been produced by analyzing the corresponding full datasets.

During the 2011 run the instantaneous luminosity of proton-proton collisions delivered by the LHC reached up to 3.5 x $10^{33}$ cm$^{-2}$ s$^{-1}$. At a bunch crossing rate of 50 ns, the average number of pp interactions per bunch crossing was approximately 10. In such conditions, the rare hard collision which produces the physics objects recognized by the trigger system and fulfilling the criteria for data acquisition –electrons, muons, taus, photons, energetic jets, missing transverse energy– is usually accompanied by several additional pp interactions overlapping with it in the same bunch crossing. These additional collisions, which are typically of low energy but may still produce significant contributions to global event characteristics such as total visible energy or charged particle multiplicity, are denoted as pile-up events. The analysis of the hard collision properly includes the effect of pile-up, which is also modeled in all the necessary Monte Carlo (MC) simulated samples.

## 3. Precision Measurements of Electroweak Observables

At a centre-of-mass energy of 7 TeV of proton-proton collisions, the LHC can be aptly described as a top quark factory: of the order of 800 thousand top quark pairs have been produced in the core of CMS in the 2011 run. Even larger is of course the number of produced W and Z bosons (respectively 500 millions and 150 millions). Although for a few selected standard model measurements –the mass of the Z boson being the clearest example– the sheer statistics of the resulting analyzable datasets cannot compensate the less clean environment of proton-proton collisions and the undetermined centre-of-mass energy of the hard subprocess with respect to the electron-positron collisions studied in the past by the SLC and LEP/LEP II colliders, a number of properties of electroweak interactions can be determined with unprecedented accuracy by the analysis of CMS data, challenging theoretical predictions. In what follows we summarize only a few of the many new measurements produced by CMS with vector bosons and top quarks.

### 3.1 Vector Boson Production Cross Sections

The production cross section of W and Z bosons at a centre-of-mass energy of 7 TeV has been studied both inclusively[14] and as a function of the number of hadronic jets accompanying the bosons[15]. Additional measurements have determined the cross section of W$\gamma$ and Z$\gamma$ production[16], as well as the production rate of WW, WZ, and ZZ pairs[17]. In all these measurements, W boson candidates are selected by searching for their decay to e$\nu$ and $\mu\nu$ final states, and Z bosons by searching for their ee and $\mu\mu$ final states. In the case of ZZ production, however, the second boson has also been identified in its decay to $\tau$-lepton pairs.

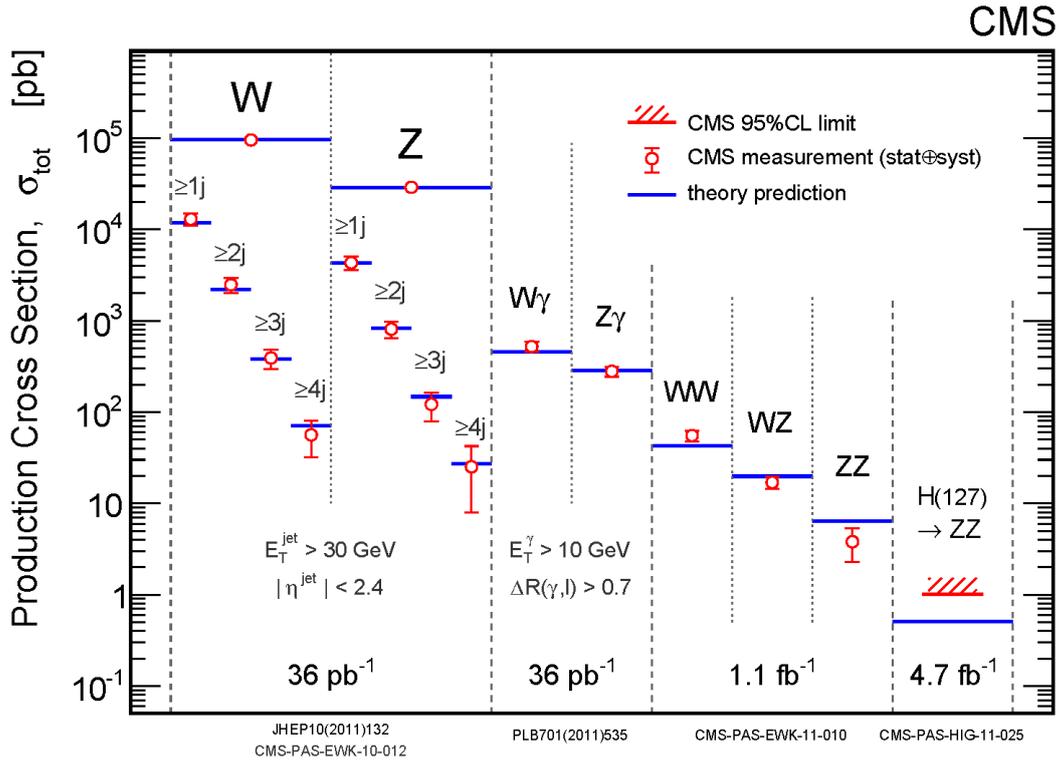

*Figure 3: The total production cross section of final states including W and Z bosons, in picobarns (red markers) are compared to theoretical predictions (blue lines). For single boson production are also reported the cross sections of processes including at least one to at least four hadronic jets with transverse energy $E_T>30$ GeV and pseudorapidity $|\eta|<2.4$. For comparison, the upper limit on the rate of decays to Z-boson pairs of a 127 GeV Higgs boson is also reported and compared to standard model predictions.*

The general picture that can be drawn is one of excellent agreement with theoretical calculations, which are available at next-to-leading order (NLO)[18-20] and next-to-next-to-leading order (NNLO)[21-25] in perturbative QCD. Figure 3 provides a very nice summary of the CMS measurements of these processes.

For some detail we may focus on inclusive W and Z production processes, which probe relatively small values of parton momenta, with fractions x in the range 0.001-0.1 for the centrally-produced vector bosons exploited by the analyses. Theoretical predictions here suffer from a significant source of uncertainty due to the imperfect knowledge of parton distribution functions (PDF). This is due in particular to the fact that vector boson production, which in hadronic interactions proceeds mainly through the Drell-Yan process, requires at the LHC a sea quark in the initial state; given the large number of gluons in the proton at the required energy scale, the scattering sea quark is mostly provided by the gluon splitting process, with the result that the uncertainty in g(x), the gluon PDF, significantly affects the calculation. Other uncertainties are due to higher-order QCD and electroweak corrections. Overall, the experimental uncertainties are still larger than theoretical ones, mainly because of the 4% error in the integrated luminosity, which is a limiting factor in all absolute cross section determinations.

The luminosity uncertainty can be eliminated by computing the cross section ratio

$$R = \frac{\sigma(W)B(W \to l\nu)}{\sigma(Z)B(Z \to ll)}$$

where B identifies the indicated branching ratio. Using electron and muon datasets from data collected in the 2010 run and corresponding to 36 inverse picobarns of integrated luminosity, CMS finds *R = 10.54 ± 0.07 (stat.) ± 0.08 (syst.) ± 0.16 (th.)*, where the first uncertainty is statistical, the second is systematic, and the third comes from theoretical uncertainties affecting detector acceptance. The NNLO prediction is $R^{th}$ = *10.74 ± 0.04*, in good agreement with the CMS result.

A measurement which is unique to proton-proton collisions is the one of the ratio between positive and negative W boson production cross section, which differs from unity due to the difference in the PDF of up and down quarks in the proton. Combining electron and muon final states CMS measures a ratio *[σ(W$^+$)B(W$^+$→lν)] / [σ(W$^-$)B(W$^-$→lν)] = 1.421 ± 0.006 (stat.) ± 0.014 (syst.) ± 0.029 (th.)*, in excellent agreement with the NNLO theoretical prediction of 1.43 ± 0.01.

### 3.2 Top Quark Mass and Cross Section Measurements

The large samples of top pairs produced in the 2011 run of the LHC have allowed the CMS experiment to measure with great accuracy the top pair production cross section in several different final states[26-28]. A summary of these determinations is shown in Fig. 4, where a comparison with the most recent theoretical predictions at NLO and NNLO[29-31] indicates an excellent agreement.

Similarly, the top quark mass has been measured by CMS both in single-lepton[32] and in dilepton final states[27]. The most recent result, which employs the full statistics of the 2011 run, is the second most precise top quark mass determination in the world at the time of writing. The analysis uses top pair decays in the "muon plus jets" topology, which arises when one of the two W bosons emitted in the $t\bar{t} \to W^+bW^-\bar{b}$ reaction decays via $W \to \mu\nu$, while the other W boson creates a pair of quarks. By selecting events from muon-triggered data which contain a clean and isolated muon candidate of $p_T>30$ *GeV* and four or more hadronic jets of $E_T>30$ *GeV*, two of them with identified secondary vertices from b-quark decay, a sample of 8094 candidates is isolated. A kinematic fit using the world average value of the W boson mass as a constraint is used to determine an estimate of the top quark mass for each possible combination of jet assignments to the final state partons. Events with no combinations passing a cut on the fit probability are discarded, resulting in a very high-purity sample of 2391 events. The estimated masses of all combinations passing the selection are finally used, weighted by their fit probability, in a global likelihood which combines information on the top quark mass and the jet energy scale factor, the latter obtained by comparing the pre-constraint mass of the jet pair assigned to the W decay with the world average W mass.

The top quark mass is measured to be *$m_t$ = 172.6 ± 0.6 ± 1.2 GeV*, where the first uncertainty quoted is the combination of statistical with jet-energy-scale-related systematic uncertainty, and the second is the quadrature sum of all other systematic uncertainties. Figure 4 shows a comparison of this estimate with other CMS determinations, along with the result of its combination with other measurements based on 2011 data.

### 3.3 Studies of Top Quark Properties

In the standard model the top quark constitutes no exception to the rule that modification of quark flavor is only caused by the charged weak current. Flavor-changing neutral current (FCNC) decays of the top quark are therefore only possible via loop diagrams, which are heavily suppressed; in particular, theoretical calculations[33] predict that the branching fraction of $t \to Zq$ decay is of the order of $10^{-14}$, and thus completely unobservable at the LHC. Even small new physics contributions to such reaction[34] could therefore be detectable in large enough data samples. CMS has searched for the mixed decay $t\bar{t} \to ZqWb$ in 2011 data, finding no significant signal and setting a 95% confidence level (C.L.) upper limit *B(t$\to$Zq)<0.34%*[35].

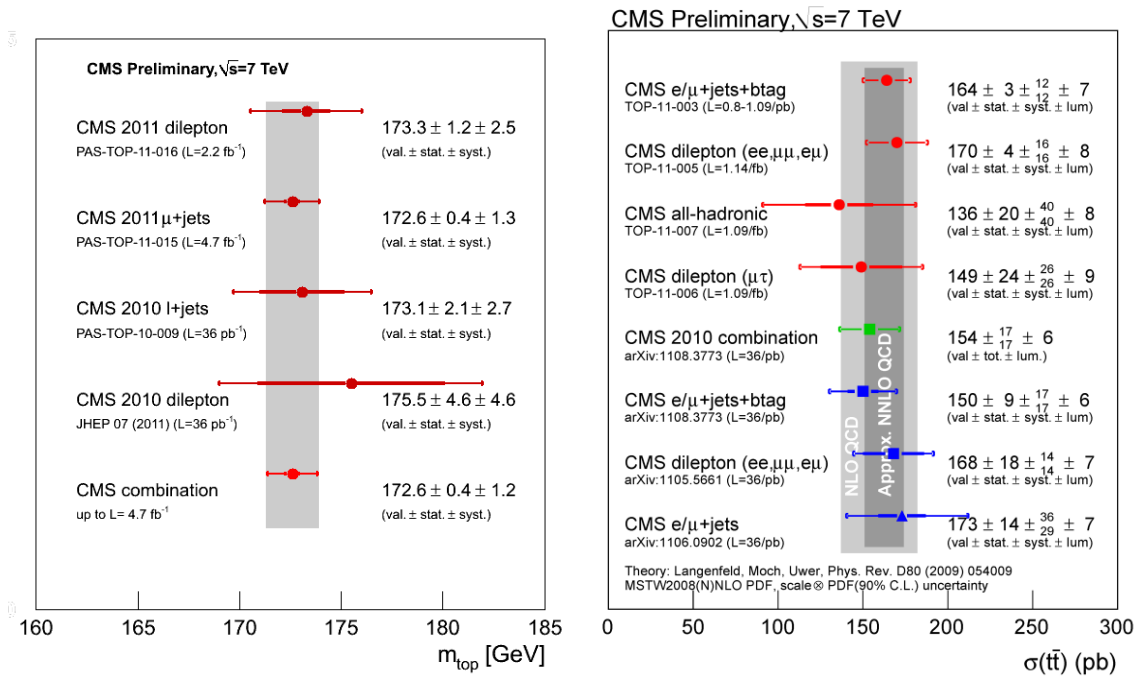

Figure 4. *Left: top quark mass determinations by the CMS experiment. From top to bottom are shown the measurements in the dilepton and µ+jets final states employing 2011 data, the analysis results of 2010 data in the single-lepton and dilepton topologies, and the combination of the four results. Right: CMS measurements of the top pair production cross section. From bottom to top are shown three results using data corresponding to 36 inverse picobarns of integrated luminosity analyzed from the 2010 run (in blue), their combination (in green), and results based on a quarter to a fifth of the total 2011 statistics (in red).*

## 4. Selected Results in Heavy Flavor Physics

In an era when heavy flavor physics was to be dominated by B factories –machines providing electron-positron collisions at the Y(4S) centre-of-mass energy– experiments studying heavy flavor properties in hadronic collisions have gone far beyond being complementary to the dedicated electron-positron experiments. In particular, the CDF experiment could match and in some cases surpass the precision of several BaBar and Belle results in the latter's own field of excellence, thanks to its innovative Silicon Vertex Tracker, a revolutionary trigger which relied on the precise online measurement of the impact parameter of charged tracks for the selection of large datasets enriched in B-hadron decays. The triggering on track impact parameter, however, is not currently possible in CMS and ATLAS, mainly because of two otherwise strong points in the design of these experiments: the one-order-of-magnitude higher bunch-crossing rate of the LHC, together with the two-orders-of-magnitude larger number of readout channels of the silicon trackers of new generation of which the LHC experiments are endowed.

The huge cross section of processes yielding bottom quarks in the final state of 7-TeV proton-proton collisions again comes to the rescue. CMS can collect semileptonic B-hadron decays of medium to large transverse momentum thanks to inclusive electron and muon triggers, as well as exploit the significant branching ratio of B-hadron decays to J/ψ and ψ(2S) mesons by triggering on the resulting pairs of very low transverse momentum muons. Indeed, the simple graph showing the invariant mass of muon pairs collected by double muon triggers speaks volumes about the heavy flavor potential of the experiment (see Fig. 5).

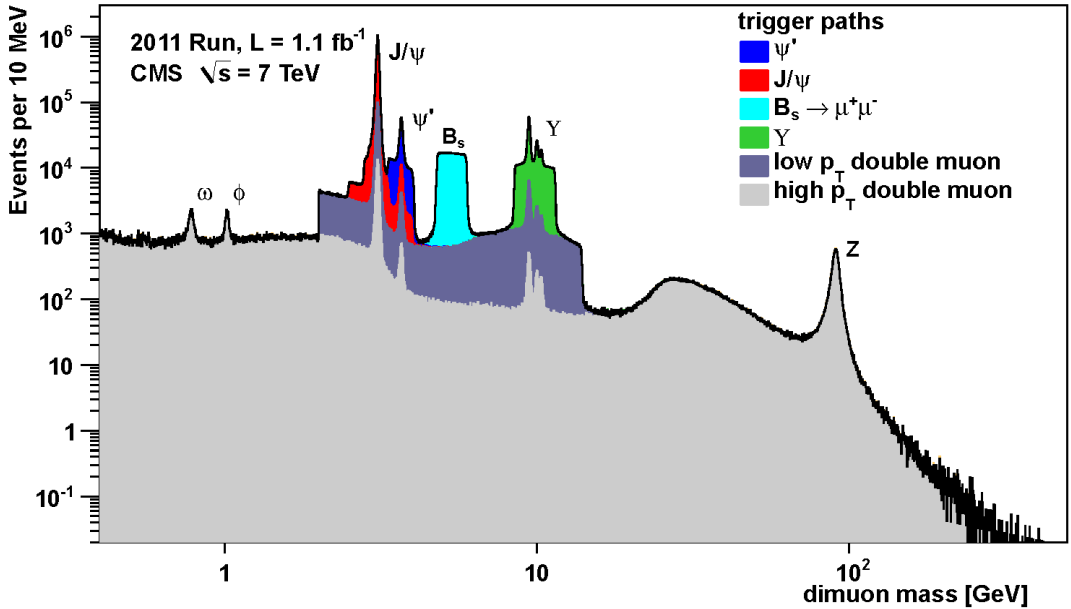

*Figure 5. Invariant mass distribution of pairs of muon candidates of opposite sign collected by muon triggers in 1.1 inverse femtobarns of 7-TeV proton-proton collision in 2011. One immediately recognizes the signals due to muon pair decays of the φ, the ω, the J/ψ and the ψ(2S), as well as the three lowest bottomonium states and the Z boson peak.*

It is therefore not surprising that CMS can now claim the most stringent limits on the rate of rare $B \to \mu\mu$ decays[5], nor that many other world-class results are being produced in this area of research. In the following we only make a brief mention of the most relevant new measurements.

## 4.1 Searches for $B_d \to \mu\mu$ and $B_s \to \mu\mu$

The search for the rare decays of neutral $B_d$ and $B_s$ mesons to muon pairs has been a hot topic in the last few years. These decays are heavily suppressed in the standard model due to the absence of flavor-changing neutral-current diagrams at tree level; two additional factors further reduce their rate: the ratio $m_\mu^2/m_B^2$ between the squared masses of muons and B mesons implied by the helicity configuration of zero-total-momentum energetic fermion-antifermion final states, and the ratio $f_B^2/m_B^2$ (where $f_B$ is the B decay constant) due to the inner annihilation of quarks in the decaying meson. The smallness of the total predicted branching fractions, $B(B_d \rightarrow \mu\mu) = (3.2 \pm 0.2)10^{-9}$ and $B(B_s \rightarrow \mu\mu) = (1.0 \pm 0.1)10^{-10}$[36], constitute an opportunity to search for indirect evidence of new physics, which could intervene in the form of the exchange of new virtual particles, with significant increases in the rate of these decays for specific values of the new physics parameters. For example, in minimal supersymmetric extensions of the standard model the decays receive enhancements for large values of tanβ, the ratio of vacuum expectation values of the two Higgs boson doublets.

---

[5] The LHCb collaboration has since updated its result[2] which is more stringent than the one reported here.

The search method is a counting experiment of events in the signal region of the dimuon mass distribution for both B species. Because of the dependence of mass resolution and background levels on the pseudorapidity of detected muons, two separate samples are analyzed independently and then combined: "barrel" candidates have both muons with $|\eta|<1.4$, and "endcap" candidates include all remaining events. MC simulations are used to estimate backgrounds from other B decays, while combinatorial backgrounds are evaluated from the data in suitable mass sidebands. A normalization sample of $B^+ \rightarrow J/\psi\ K^+$ decays, with the subsequent $J/\psi \rightarrow \mu\mu$ decay, is collected by the same dimuon trigger used as a starting point of the rare decay search, and is used to remove the uncertainties of B hadron production cross section and integrated luminosity of the data sample.

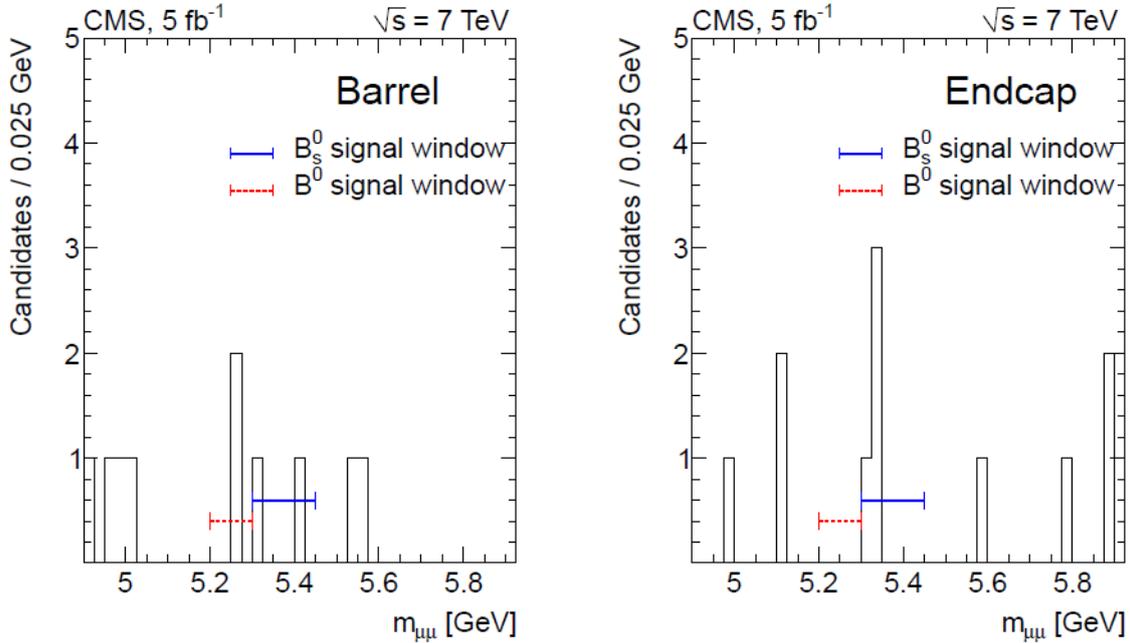

Figure 6. *Reconstructed invariant mass of muon pairs for $B_s$ and $B_d$ candidates selected by the CMS search in the total 2011 dataset. The data is divided in muon pairs identified in the central region (Barrel, left) and in the forward region (Endcap, right). The search windows for $B_s$ and $B_d$ mesons are displayed as horizontal blue and red bars, respectively.*

Muon candidates are selected using the quality of their track fits. Muon pairs with invariant mass in the *4.9<$M_{\mu\mu}$<5.9 GeV* range are then preselected, and candidates in the signal regions (*5.2<$M_{\mu\mu}$<5.3 GeV* for the $B_d$ and *5.3<$M_{\mu\mu}$<5.45 GeV* for the $B_s$) are blinded to avoid potential bias in the selection procedure. A random-grid search of 1.6 million possible selection strategies is performed on a set of variables capable of discriminating the rare B decay signals from backgrounds, optimizing the sensitivity to the expected upper limit; among the used variables are the $\chi^2$ of the fit of muon trajectories to a common vertex, the transverse momentum of the higher-$p_T$ and the lower-$p_T$ muon, the B candidate transverse momentum, the isolation of the muons from other tracks in the event, the number of nearby tracks, and the smallest impact parameter of these tracks with respect to the common vertex of the muon pair.

      Figure 6 shows the mass distribution of final candidates accepted by the optimized selection. In the "barrel" sample two candidates are found for each B meson species, while in the "endcap" four $B_s$ candidates and zero $B_d$ candidates are observed. Upper limits on the branching ratios are determined at 95% C.L. using the $CL_s$ criterion[37-38]: for the $B_s$ meson the limit is $B(B_s \to \mu\mu) < 7.7 \; 10^{-9}$, and for the $B_d$ meson the limit is $B(B_d \to \mu\mu) < 1.8 \; 10^{-9}$. These limits can be used to constrain several proposed extensions of the standard model.

### 4.2 B Cross Section Measurements

The study of production cross sections of b-quarks in hadron collisions enables detailed tests of QCD calculations at next-to-leading order, which suffer from sizable uncertainties due to the choice of renormalization and factorization scales. Comparisons between experimental measurements and the results of calculations employing perturbative expansions in powers of leading logarithms, *e.g. $ln(p_T/m_b)$* which are important at high transverse momentum $p_T$ due to multiple gluon radiation processes, are useful to test the validity of those approaches. In the past, cross section determinations at lower energy, *e.g.* those produced at the HERA[39-40] and Tevatron[41-42] colliders, showed intriguing disagreement with theoretical predictions; the sources of the discrepancies have been largely understood[43-45], but it remains important to extend the investigations at the higher centre-of-mass energy provided by LHC collisions, where QCD is tested in an extended kinematical domain.

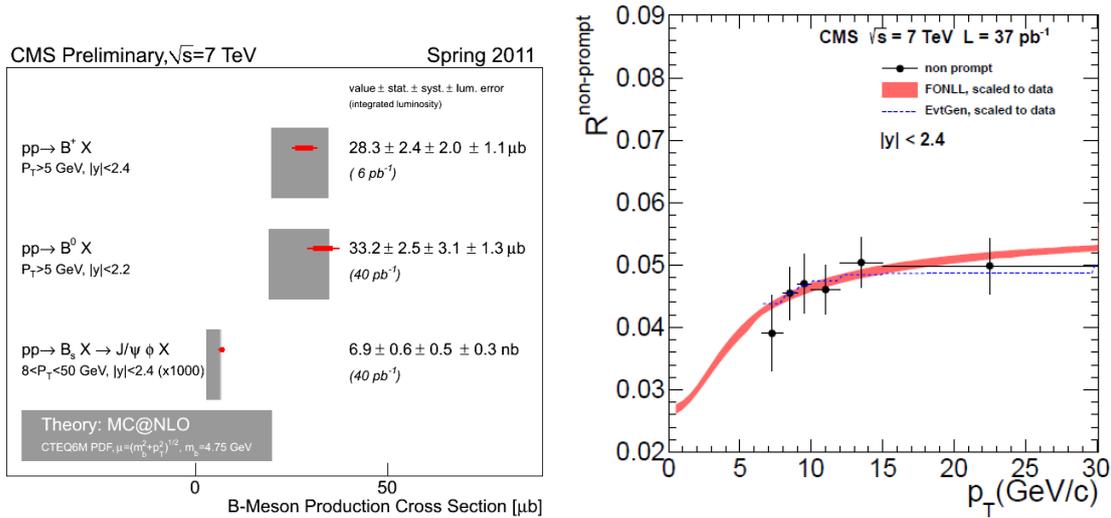

Figure 7. *Left: comparison of different determinations of the B hadron cross section in 7-TeV proton-proton collisions (red points with thick statistical and thin systematic error bars) with MC@NLO calculations. Right: the ratio between non-prompt ψ(2S) and J/ψ production measured as a function of meson transverse momentum is compared with the $p_T$ shape of a fixed order next-to-leading-log (FONLL) calculation.*

CMS produced several measurements of B hadron cross sections already with the relatively small dataset collected during 2010, using a variety of technologies[46-50]; a summary is provided in the left panel of Fig. 7, which compares those results with MC@NLO calculations[51] using the CTEQ6M set of parton distribution functions[52]. The agreement is overall good. The analysis of the much larger 2011 dataset will allow differential measurements, with more stringent tests of the theory.

As part of a study of J/ψ and ψ(2S) production[53], CMS has also recently produced a measurement of the ratio between the fraction of B hadrons decaying to a ψ(2S) meson and the fraction decaying to a J/ψ. The measurement has been produced as a function of the meson $p_T$ and rapidity; no significant dependence on rapidity has been observed, so CMS reported the ratio as a function of $p_T$ only (Fig. 7, right), which agrees with the prediction of a fixed-order next-to-leading-logarithm calculation. A further result of that study is the determination of the inclusive branching ratio of B hadrons to ψ(2S) mesons, which is measured as *B(B→ψ(2S)X) = (3.08 ± 0.12 ± 0.13 ± 0.42)10⁻³*, where the first uncertainty is the combination of statistical and systematic effects, the second is due to residual theoretical uncertainties, and the third is due to the uncertainty in the world average branching fractions of *B→J/ψX, J/ψ→μμ*, and *ψ(2S)→μμ* decays. The result is in good agreement with the world average value of *(4.8 ± 2.4)10⁻³*.

## 5. Searches for the Higgs Boson

Probably the most impatiently awaited results of the LHC experiments in 2011 have been the ones of searches for the standard model Higgs boson[54-59]. Predictions produced at the end of the 2010 run (see Fig. 8) implied that the analysis of 5 inverse femtobarns of 7-TeV proton-proton collisions would allow the 95% C.L. exclusion of the particle for all mass hypotheses up to 600 GeV, if the particle did not exist in that mass range, or the detection of signals whose significance varied from the level of a conclusive observation to that of a tantalizing hint. On December 13$^{th}$ 2011 the ATLAS and CMS collaborations released preliminary results of those searches which indeed fulfilled the predictions: both experiments claimed a 95% C.L. exclusion of the Higgs hypothesis in most of the searched mass range, and a first evidence, consistent across the two experiments albeit still inconclusive, of a possible 124-125 GeV signal. In this section we provide a review of the CMS results, which in the meantime have consolidated into several publications[60-68].

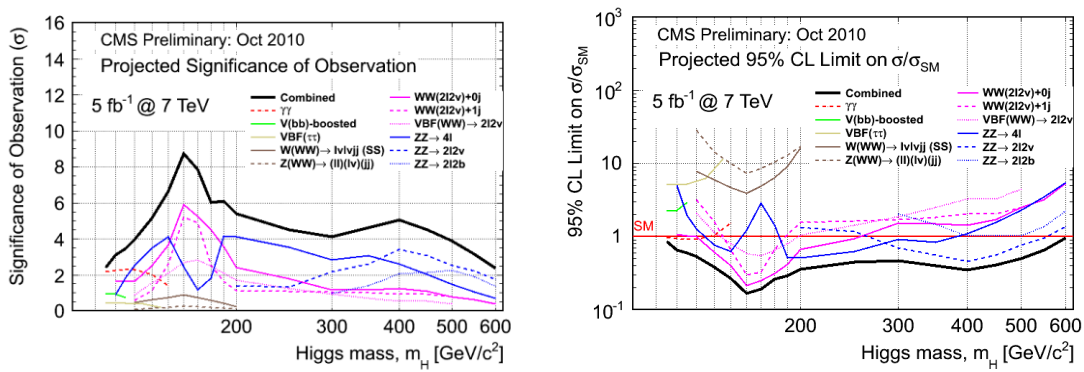

Figure 8. *Predicted significance of a Higgs boson signal (left) and 95% C.L. upper limit on the signal cross section in SM prediction units (right) resulting from the combined CMS analysis of all sensitive search channels using 5 inverse femtobarns of 7-TeV proton-proton collisions.*

### 5.1 Production and Decay

The standard model Higgs boson has a non-zero coupling to all massive particles, and can therefore be produced by several different mechanisms in proton-proton collisions. From an experimental standpoint the most important reactions at the LHC include gluon fusion diagrams, where a Higgs boson is emitted most frequently by a virtual top-quark loop; vector-boson-fusion processes, where the Higgs is produced together with two characteristic high-rapidity hadronic jets resulting from the emission of two virtual W or Z bosons off the initial state quarks; and Higgs-strahlung diagrams, where the particle is radiated by a highly-off-shell W or Z boson. A graph of the cross sections predicted for these processes as a function of the unknown value of the Higgs boson mass is given in Fig. 9 (left).

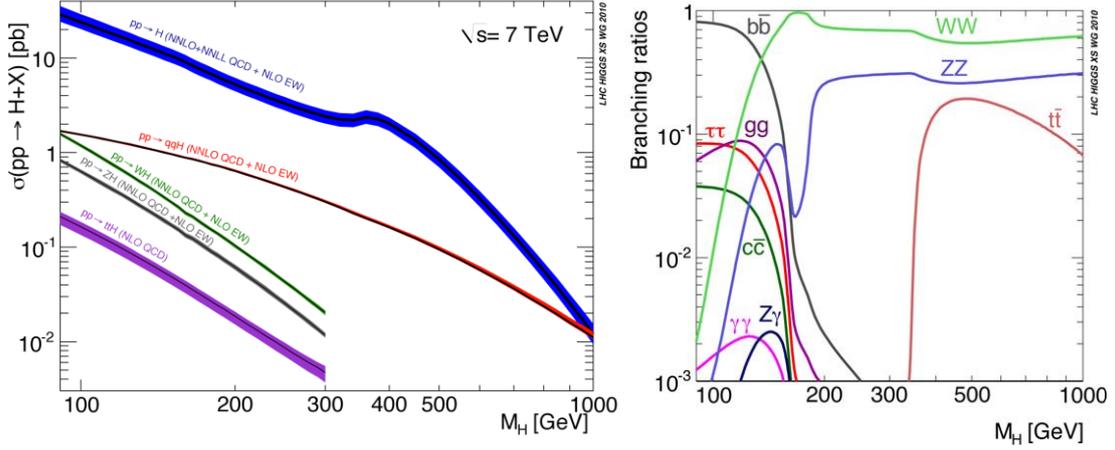

Figure 9. *Left: predicted cross section of Higgs production processes in 7 TeV proton-proton collisions as a function of Higgs boson mass. From top to bottom the curves describe gluon-fusion, vector-boson fusion, associated WH and ZH production, and Higgs-strahlung off a top quark pair. Right: Higgs boson branching fractions as a function of Higgs boson mass.*

The Higgs mass is also the crucial parameter in determining the expected admixture of decay modes (see Fig. 9, right). One typically distinguishes between a "low-mass" range ($m_H<135\ GeV$) where the dominant decay is $H \rightarrow b\bar{b}$, and a "high-mass" range ($m_H>135\ GeV$) where the decay produces most frequently pairs of W or Z bosons. However, a significant branching fraction is retained in the low-mass range by $WW^*$ and $ZZ^*$ states, where one of the bosons is off mass-shell. From an experimental standpoint this has important consequences; in particular, a 125 GeV Higgs boson ends up being visible in a large number of different final states: $H \rightarrow b\bar{b}$, $H \rightarrow \gamma\gamma$, $H \rightarrow \tau\tau$, $H \rightarrow ZZ^*$, and $H \rightarrow WW^*$. While the decay to b-quark pairs, which can only be sought for in Higgs-strahlung processes such as $W^* \rightarrow WH \rightarrow l\nu b\bar{b}$, is very difficult to extract at the LHC because of the large and irreducible W+jets background, the decay to W and Z pairs remains a sensitive probe down to the LEP II lower mass limit, 114.4 GeV[69].

## 5.2 Nuts and Bolts of the Statistical Analyses

All CMS searches of the Higgs boson, as well as their combination, employ the same frequentist technique to determine upper limits on the production cross section of the searched processes. The so-called $CL_s$ ratio[37-38] between the p-value of the signal-plus-background hypothesis $H_1$ and the p-value of the background-only hypothesis $H_0$ is used as a criterion to determine how the data conform to the two alternatives.

The details can be described considering a generic counting experiment. A likelihood function can be written in terms of the expected background and signal of the two hypotheses and a signal strength modifier μ describing the ratio between a given observed cross section σ and the standard model prediction, $\mu=\sigma/\sigma_{SM}$. Denoting by $\theta$ the vector of nuisance parameters affecting predicted signal $s(\theta)$ and background $b(\theta)$ counts, one may write

$$L\ (data\ |\ \mu,\theta) = Poisson\ [data\ |\ \mu s(\theta)+b(\theta)]\ \Pi(\theta)$$

where $\Pi(\theta)$ indicates a prior knowledge of the nuisance parameters, coming from real or hypothetical auxiliary measurements. One may then define a profile likelihood-ratio test statistics as

$$\widetilde{q}_\mu = -2\ln\frac{L(data\ |\ \mu,\hat{\hat{\theta}}_\mu)}{L(data\ |\ \hat{\mu},\hat{\theta})}.$$

In the expression above the hatted symbols on the parameters indicates that they are the respective maximum likelihood estimates (MLE), such that at the denominator one has the global maximum of L, while at the numerator appears the maximum value of L for a given value of the signal strength modifier µ; the MLE of µ at the denominator in the above expression is constrained in the range *[0,µ]*. This avoids negative signal strength solutions, and ensures that best-fit values of the signal strength above the standard model prediction are not counted as evidence against it.

Using the above definition of the test statistics, MLE of the nuisances $\theta_\mu$ and $\theta_0$ are computed for the two hypotheses, given the data. These values are used to generate pseudo-data for $H_1$ and $H_0$ and construct the respective probability density functions $f(q_\mu)$ of the test statistics, which yield p-values of the two hypotheses when integrated from the observed values $q_\mu$ and $q_0$ to infinity. One thereby constructs the ratio

$$CL_s = \frac{\int_{q_\mu^{obs}}^{\infty} f(\widetilde{q}_\mu\ |\ \mu,\hat{\theta}_\mu^{obs})d\widetilde{q}_\mu}{\int_{q_0^{obs}}^{\infty} f(\widetilde{q}_\mu\ |\ 0,\hat{\theta}_\mu^{obs})d\widetilde{q}_\mu}.$$

Values of the signal strength modifier µ for which *$CL_s$<0.05* are excluded by a 95% C.L. test. The procedure can be repeated for different Higgs mass hypotheses; if the excluded range of µ values includes the standard model prediction *(µ=1)*, the corresponding mass hypothesis is excluded, at the stated confidence level. More detail on the technique can be found in a combined ATLAS-CMS document[70].

### 5.3 Searches for H→ZZ Decays

CMS has searched for *H→ZZ* decays in the full 2011 dataset of proton-proton collisions using final states including one Z boson decay to electron-positron or muon pairs, which provides an easy triggering strategy and a very clean signature, and then selecting candidate decays of a second Z boson to any charged lepton pair (ee, µµ, and ττ), neutrinos, or hadronic jets. All of the searches contribute to test the Higgs boson hypothesis in the high-mass range *$m_H$>200 GeV*, while sensitivity to lower values of $m_H$ is provided only by the *H→ZZ→4l (l=e,µ)* and the *H→ZZ→lljj* (j=hadronic jet) signatures. A sample illustration of mass spectra resulting from these searches is provided in Fig. 10, while Fig. 11 (left) shows the upper limit on the signal strength modifier obtained from the search of ll𝜈𝜈 events. The right panel in Fig. 11 shows the event display of a nice H→ZZ→llττ candidate.

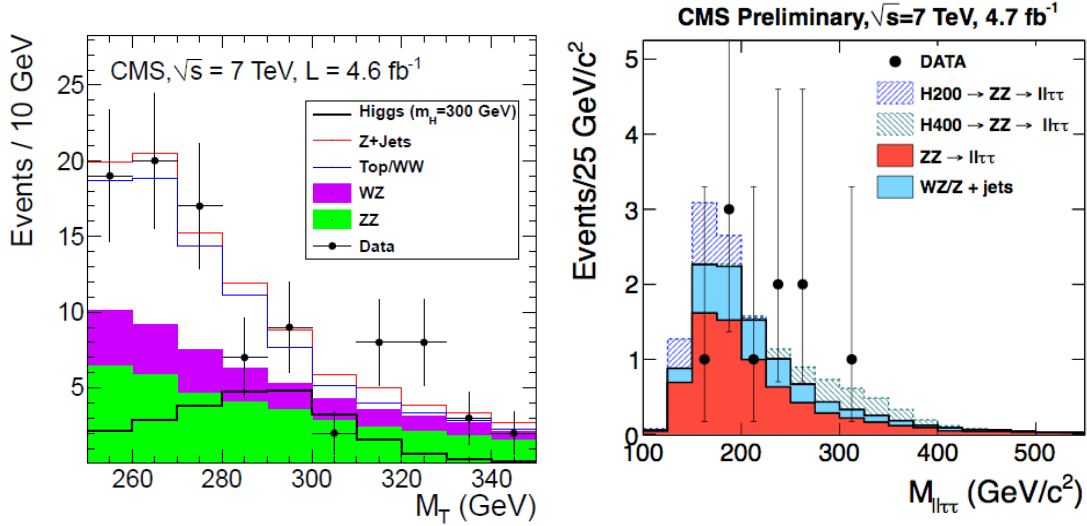

Figure 10. *Left: four-body transverse mass distribution of H→ZZ→llνν candidates. The neutrino transverse momenta are inferred from the observed transverse energy imbalance and its attribution to the escaping neutrinos in the Z→νν decay hypothesis. The data (black points) are compared to the stacked sum of expected backgrounds (colored histograms); the hypothetical contribution from a $M_H$=300 GeV Higgs boson is shown in black, unstacked. Right: reconstructed mass of llττ candidates in the H→ZZ→llττ search. Backgrounds from standard model diboson processes (ZZ, in red, and WZ or Z+jet processes, in cyan) are compared to the mass distribution of 10 observed candidates (black points). The expected Higgs signals for $m_H$=200 and 400 GeV are shown stacked as hatched histograms.*

In the golden "four-lepton" channel characterized by the eeee, eeμμ, or μμμμ final state the small backgrounds are due to standard model ZZ production, with smaller contributions from Z+bb, top pair production, and Z+jets production; all but ZZ production contribute significantly only for four-lepton masses below 200 GeV. The ZZ background is estimated from MC simulation, while other backgrounds are estimated from orthogonal selections enriched in the various background components. In the data, 72 candidates are observed, with a total background prediction of 67.1 ± 6.0 events.

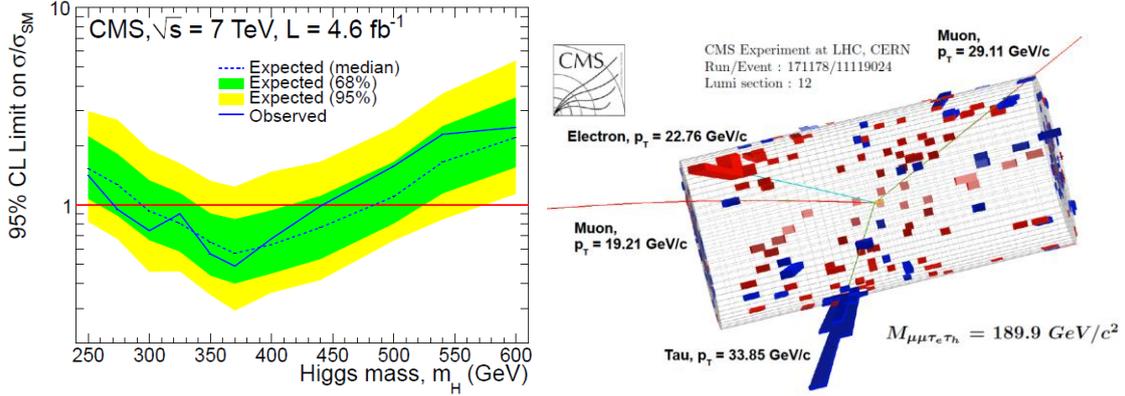

Figure 11. *Left: upper limit on the signal strength modifier µ resulting from the search of H→ZZ→llνν candidates (black curve). The green and yellow bands illustrate the 1σ and 2σ range of expected upper limits, respectively. Right: event display of a H→ZZ→µµττ candidate. The red and blue boxes, of size proportional to measured transverse energy, show the detected signals in the electromagnetic and hadronic calorimeters, respectively; tracks in red show the reconstructed trajectory of identified muon candidates, and the cyan track is associated with an electron candidate from leptonic τ decay.*

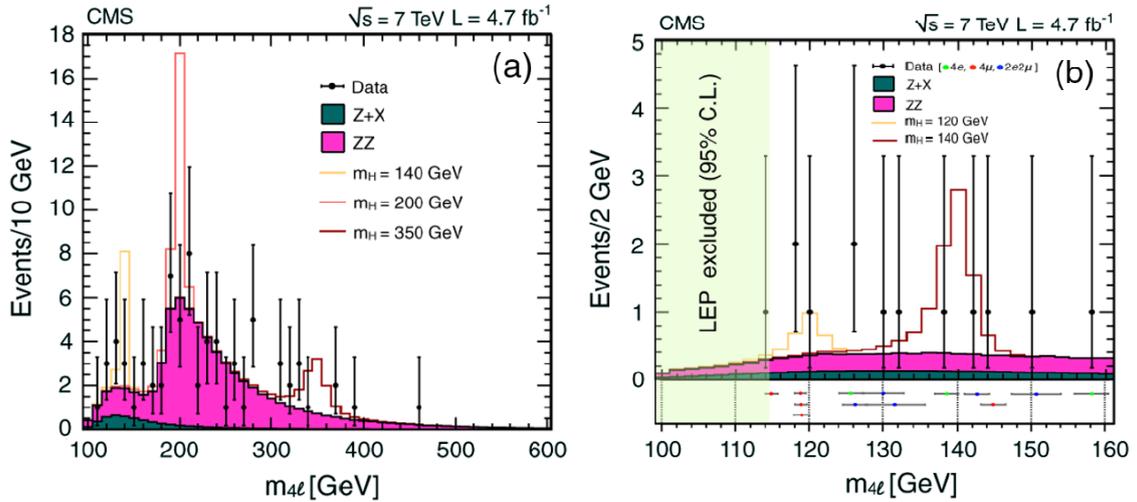

Figure 12. *Reconstructed four-lepton mass distribution of H→ZZ→eeee, H→ZZ→eeµµ, and H→ZZ→µµµµ candidates in 2011 CMS data (black points with error bars). Empty histograms describing several different Higgs mass hypotheses are stacked on top of predicted backgrounds from ZZ production (magenta) and other sources (in grey) for the sake of comparison. The left panel shows the full mass range considered in the search; the right panel shows a blow-up of the $m_H$<160 GeV region.*

The search exploits the narrow mass resolution of the four-lepton combinations through unbinned likelihood fits of the mass distribution in the data (Fig. 12) to the sum of background and a set of narrowly-spaced mass hypotheses in the 110-600 GeV range.

Slight excesses are observed for masses near 119 GeV and 320 GeV, with local significances of 2.5σ and 2.0σ, respectively. After accounting for the multiplicity of independent search regions in the mass spectrum[71], these however get reduced to values below one standard deviation. A graph of the p-value of the background-only hypothesis as a function of $m_H$ is provided in Fig. 13 (left), while the right panel shows the derived upper limits on the signal strength modifier μ versus Higgs mass.

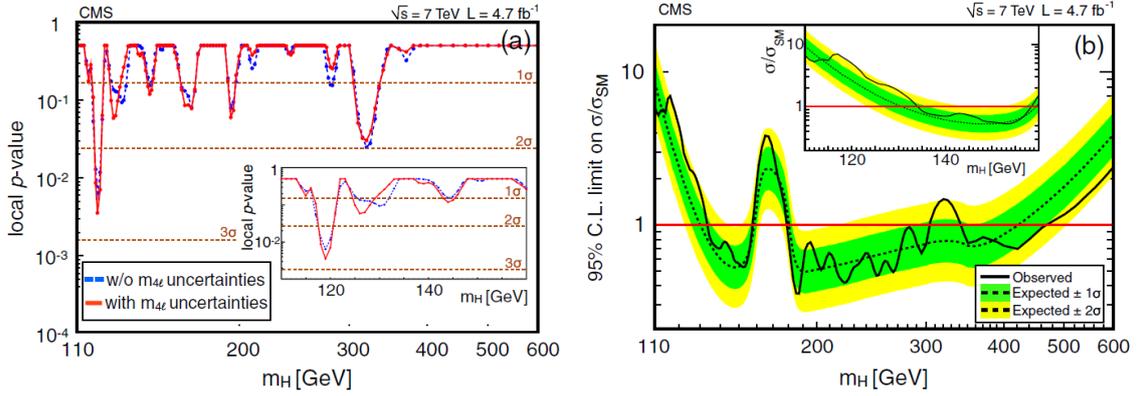

Figure 13. *Left: local p-value of the background-only hypothesis in the H→ZZ→4l search (l=e,μ) as a function of the four-lepton invariant mass; the lower-right inset shows a blow-up of the low-mass region. Right: upper 95% C.L. limit on the signal strength modifier μ as a function of Higgs boson mass. Mass hypotheses corresponding to values below 1.0 of the black curve are excluded by the search results. The green and yellow bands indicate 1σ and 2σ range of expected upper limits given the data size and the analysis method.*

### 5.4 Searches for the H→WW Decay

The *H→WW* decay mode is the most frequent one for *$m_H$>135 GeV*, and its search provides high sensitivity to a standard model Higgs boson, particularly in the range of masses around *$2m_W$*. The searched final state is the one produced by a decay to eν or μν pairs of both W bosons; the presence of two energetic neutrinos prevents the direct reconstruction of the Higgs mass. Events with two electrons or muons of opposite charge and significant missing transverse energy are divided in three categories depending on the number (0, 1, or 2) of hadronic jets of *$E_T$>30 GeV* they contain in addition to the high-$p_T$ leptons. Events with more than two jets are rejected. After the application of a tuned selection aimed at removing reducible backgrounds from Drell-Yan, WZ/ZZ, and top pair production using the kinematical characteristics of the signal, the data is dominated by real WW production events. Two alternative sets of Higgs-mass-optimized selections, one based on one-dimensional cuts and the other using multi-variate discriminants employing the boosted-decision trees (BDT) technique, are finally applied to the 0- and 1-jet selections, while only a cut-based selection is applied to the limited-sensitivity 2-jet subsample.

Backgrounds surviving the selections are largely estimated with control samples of data, except for the contribution of a few electroweak reactions (Wγ* and some residual diboson processes) which are estimated with corresponding MC simulations. Besides standard model WW production, the main contributions are due to top pair production, W+jets production where jets produce a fake lepton signal. The comparison of event counts in the data with predicted backgrounds evidences no significant signal (see e.g. Fig. 14), and allows to determine upper limits on the signal-strength modifier $\mu=\sigma/\sigma_{SM}$. Using the *a priori* more sensitive BDT selection, CMS excludes at 95% C.L. the mass region *129<$m_H$<270 GeV* using this search channel alone.

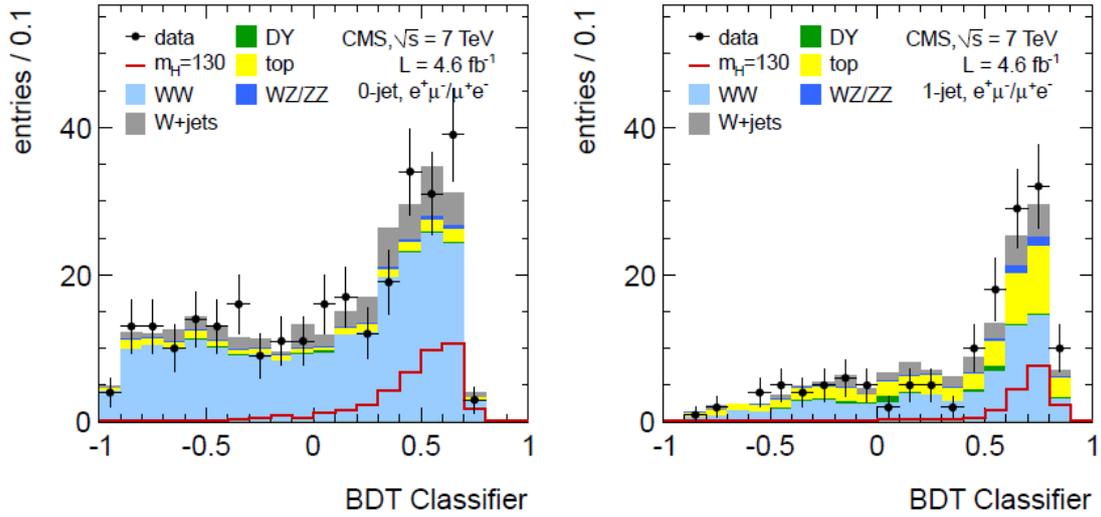

Figure 14. *BDT output in the $m_H$=130 GeV selection for the eμ final state. Left: 0-jet category; right: 1-jet category. The data (black points with error bars) are compared to the sum of predicted backgrounds, dominated by WW production (in cyan). The expected H→WW signal contribution for $m_H$=130 GeV is shown as a red empty histogram, unstacked.*

### 5.5 Searches for the H→γγ Decay

Despite the smallness of its branching fraction (approximately equal to 0.2% in the 120-130 GeV mass range), the *H→γγ* decay channel is one of the main tools for a light Higgs boson search at the LHC. The efficient and clean identification of energetic photons and their precise energy measurement are among the strongholds of the CMS detector design, with the precision of its electromagnetic calorimeter and the redundancy of the silicon tracking system.

The general strategy of the analysis is to select photon-pair candidates in various categories divided according to their background contamination and expected resolution in the two-photon-mass $m_{\gamma\gamma}$, and search for a narrow bump in the resulting $m_{\gamma\gamma}$ distributions. Particular care is put in the determination of the hard interaction vertex: the average number of pile-up interactions in 2011 collisions is about 10, and their root-mean-square separation along the beam axis amounts to 6 cm, while it is necessary to know the vertex location to ~1 cm in order to reduce to a negligible level the smearing of $m_{\gamma\gamma}$ due to the resulting uncertainty in the opening angle of the photon pair. The most probable vertex is identified with better than 80% efficiency using the kinematical properties of the observed charged tracks of each vertex as inputs to a BDT discriminant. Photon isolation is carefully exploited to increase as much as possible the separation of real prompt photon pairs from reducible backgrounds coming from γ-jet and jet-jet production.

Four independent event classes are defined based on the pseudorapidity of the most forward photon and the isolation of the most isolated one, while a fifth class collects candidates of vector-boson-fusion production of $H \rightarrow \gamma\gamma$, identified by the presence of two forward jets. The five mass distributions are separately fit to the sum of a background model, obtained from the study of the full 100-180 GeV distribution of $m_{\gamma\gamma}$, and a signal model derived from the NLO matrix-element generator POWHEG[72-73] interfaced with the PYTHIA Monte Carlo[74]. The fit results are combined to extract a global limit on the signal strength modifier μ using the $CL_s$ technique described in section 5.2. The results are shown in Fig. 15 (left). An excess of events with a local significance of 3.1σ is observed for a Higgs mass around 124 GeV.

**5.6 Other Searches for the Higgs Boson**

Besides the three main search channels discussed above, all characterized by the decay to pairs of vector bosons, CMS has produced results of two fermionic decay modes of the standard model Higgs boson: b-quark pairs and tau-lepton pairs. These are less sensitive to a standard model Higgs than the bosonic channels, but they provide valuable input to the combined search.

The $H \rightarrow b\bar{b}$ decay is sought for in combination with the signal of a highly-boosted W (Z) boson decaying to eν or μν (ee, μμ, or νν) pairs. The neutrino final state of Z bosons is flagged by the presence of large missing energy pointing away from any jet activity in the transverse plane; two b-tagged jets characterize the Higgs decay signature. Backgrounds arise from vector boson plus jet production, top pairs, and QCD multijet production; their contributions are measured in suitable control regions. Two separate searches are performed, one using a cut-based selection on the dijet mass of the b-tagged jets, the other employing a BDT discriminant. Upper limits are set on the strength modifier μ at values ranging from approximately three to ten, for Higgs masses in the 110-135 GeV range.

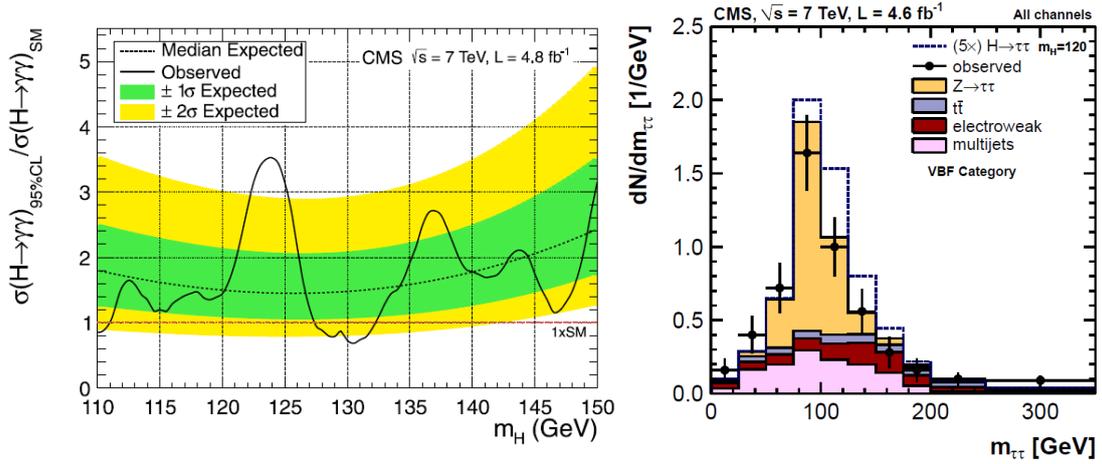

Figure 15. *Left: upper 95% C.L. limit on the signal strength modifier μ as a function of Higgs boson mass from the analysis of H→γγ signatures[6]. Right: reconstructed ττ mass distribution of τ pairs in the vector-boson-fusion search for H→ττ decays. The orange histogram represent the estimated Drell-Yan contribution, while the empty white histogram shows the contribution of a Higgs decay with μ=5.*

In the H→ττ search backgrounds from Drell-Yan produced τ pairs overwhelms the small expected signal of direct Higgs production, therefore two methods are used to increase the signal-to-noise ratio. One is to rely on vector-boson fusion production of the Higgs boson, the other is to require the presence of a high-$p_T$ jet in addition to the τ pair candidate. A maximum likelihood technique is used to reconstruct the most likely mass of the reconstructed τ pair system, thus partly recovering the missing information from the neutrino(s) emitted in τ decay. The mass distribution of candidates in the BDT search category is shown as an example in Fig. 15 (right), where the Z→ττ contribution is clearly visible. Upper limits on the signal strength modifier μ are obtained at values varying from 3 to 7 for $m_H$ ranging from 110 to 145 GeV.

### 5.7 Combined Results

The results of the five searches described above are combined by CMS by taking into account all statistical and systematic uncertainties and their correlations[76]. There are a total of 42 exclusive sub-channels that contribute to the combined results, with a total of 156 to 222 individual sources of systematic uncertainty, depending on 183 different Higgs boson mass hypotheses in the 110-600 GeV range; the interval between considered mass hypotheses is driven by the expected mass resolution at low $m_H$, and by the intrinsic Higgs boson width at large $m_H$. The combination is performed by constructing a global likelihood function, with each systematic source assigned to a nuisance parameter; each of these has a corresponding probability density function, and most of these are constrained by subsidiary measurements. The method is the one already explained in Sec. 5.2 above.

---

[6] A newer result is available for this analysis[75].

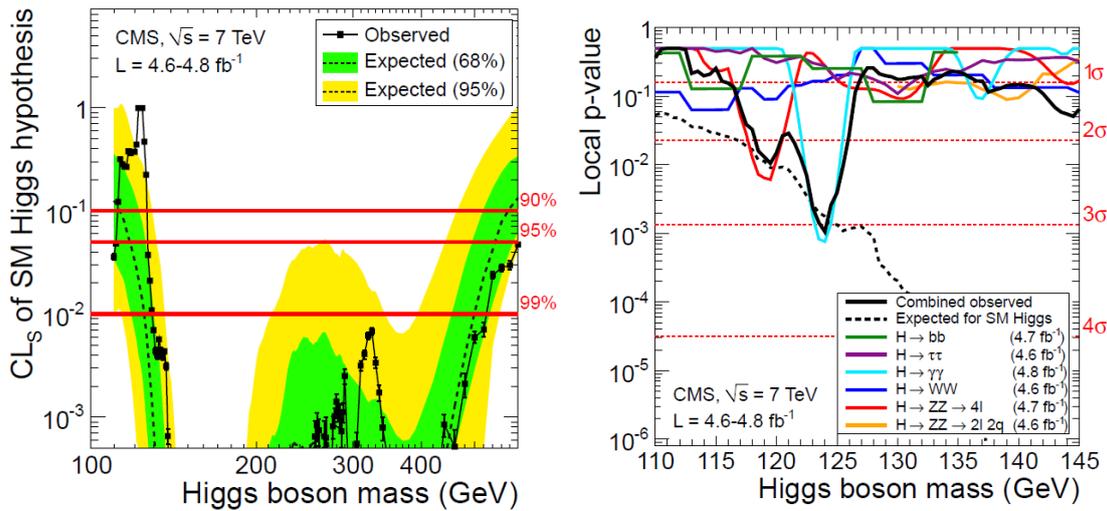

Figure 16. *Left: CL$_s$ value of the combined CMS Higgs boson search from 2011 data as a function of the hypothesized Higgs boson mass. Right: local p-value of the background-only hypothesis as a function of m$_H$. Different search channels are indicated by different colored curves; the p-value predicted in the presence of a real Higgs boson is indicated with a dashed curve.*

The CL$_s$ of the SM Higgs hypothesis is shown as a function of Higgs mass in Fig. 16. This allows CMS to exclude the range *127<m$_H$<600 GeV* at a confidence level of 95%.

The right panel in Fig. 16 shows the local p-value of observed excesses in the data, as a function of m$_H$. For *m$_H$=124 GeV* an excess corresponding to a 3.1σ significance is observed[7], but the significance of the observation is reduced to 1.5σ (2.1σ) if the multiplicity of mass hypotheses in the full 110-600 GeV range (the 110-145 GeV range) is considered.

Finally, Fig. 17 shows the best-fit value of μ in the low-mass region from the full set of searches as a function of m$_H$. The data is globally compatible with the predicted standard model signal strength (μ=1) in the region 117-126 GeV. For the hypothesis m$_H$ =124 GeV the best-fit value is also shown for each individual channel, showing that there is overall compatibility in the results. If this signal is indeed due to standard model Higgs boson production, 8-TeV data from the 2012 run of the LHC are very likely to certain to ascertain it.

---

[7] All combined Higgs search results have been updated by a recent CMS publication[1]. The highest local significance is there reported as a 2.8σ effect for *m$_H$=125 GeV*, and the corresponding corrected significance is 0.8σ.

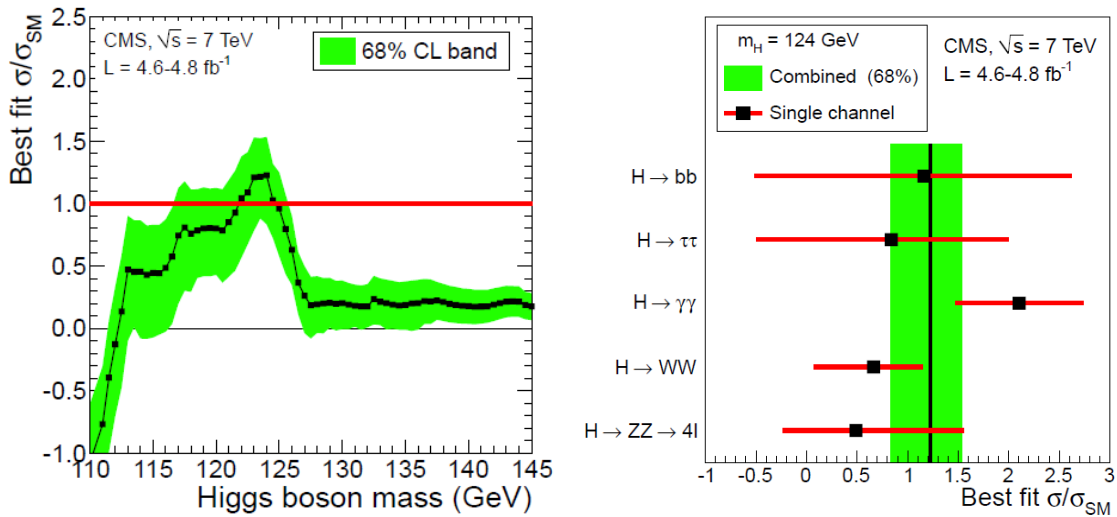

Figure 17. *Left: best-fit value of the signal strength modifier μ as a function of Higgs mass in the low-mass range 110<$m_H$<145 GeV. The band includes statistical and systematic uncertainties. Right: for the $m_H$=124 GeV hypothesis the best-fit value of μ is shown individually for the five analyzed decay channels.*

## 6. Searches for Supersymmetry Signatures

Among all hypothesized extensions of the standard model Supersymmetry is one of the most studied alternatives. The symmetry between standard model fermions and bosons with supersymmetric counterparts of bosonic and fermionic nature automatically cancels the large quantum contributions to the Higgs boson mass due to virtual loops of standard model particles[77-80], solving the naturalness puzzle in a very elegant way; the added bonus is a unification-ready merging of coupling constants below the Planck scale. Supersymmetric particle searches have been carried out in the past thirty years without success, pushing the allowed mass of the hypothetical SUSY particles to higher and higher values. Despite those early results, the much larger centre-of-mass energy of the LHC led many to trust that SUSY particles would suddenly pop up soon after the start of data taking, with unmistakable and striking signatures. But Nature has chosen otherwise.

The CMS experiment has searched the 2011 data for supersymmetric particle signatures in a number of final states and with a variety of advanced methods[81-91]. Here, for the sake of brevity, only a summary of those searches is provided. In general, SUSY particles can be copiously produced in LHC proton-proton collisions in the form of pairs of squarks or gluinos, which carry color quantum numbers and are thus subject to strong interactions. Depending on the mass spectrum of SUSY particles, the decay of squarks and gluinos may give rise to several lighter supersymmetric states in succession, with a typical "cascade" signature and characteristic kinematic features. At the end of the decay chain, a quite general signature of R-parity-conserving SUSY theories is the production of neutral weakly-interacting particles called neutralinos, which are the lightest in the supersymmetric spectrum and are thus stable. Their escape from the detector with large transverse momentum can be flagged by the same experimental observable used to detect neutrinos, *i.e.* a large energy imbalance in the plane transverse to the beams. Experimental searches often require large values of missing transverse energy, in some cases along with hadronic jets, in others accompanied with charged leptons or more complex final states.

Figure 18 summarizes the status of CMS searches for SUSY particles, in 2011 datasets corresponding to up to 2.1 inverse femtobarns of integrated luminosity. Upper limits in production cross sections are turned into exclusion regions in the plane described by the universal scalar and gaugino masses. The most sensitive searches are the one for jets and missing energy and the "razor" analysis, which exploits the kinematic configurations of the jets in the reconstructed reference frame of super-particle decay. Those searches are expected to produce much tighter limits on superparticle masses when performed on the data from the 2012 run, because of the larger statistics as well as the significant increase of the production cross section for very massive objects.

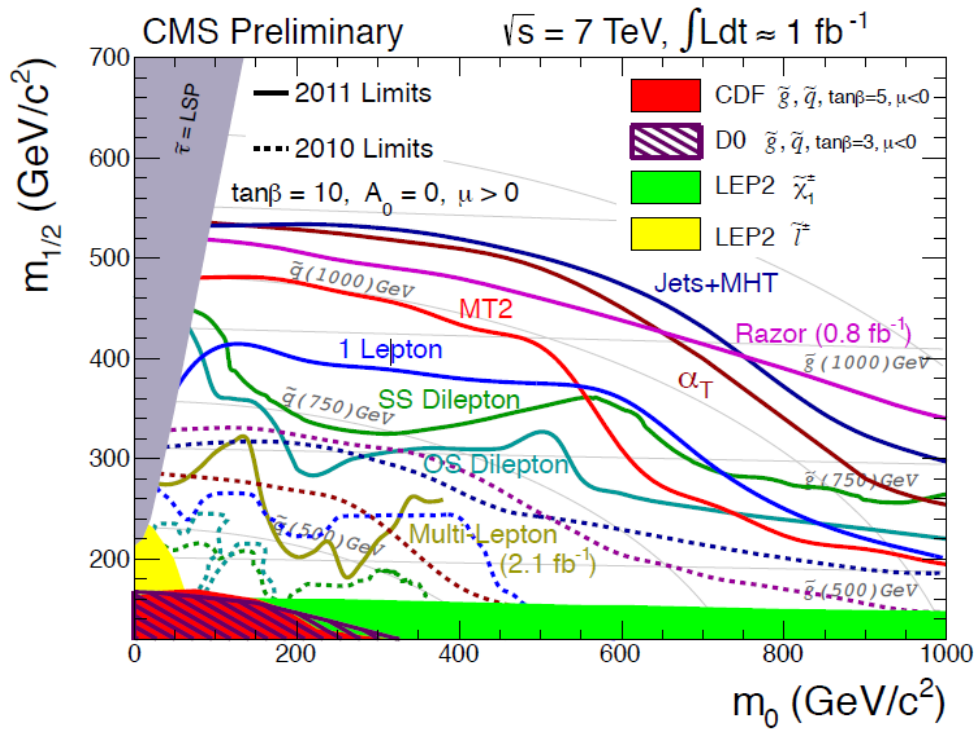

Figure 18. *Summary of results of SUSY searches by CMS using 2010 (dashed curves) and 2011 data (full curves), here shown for a representative choice of SUSY parameters. Grey curves show iso-contours in the value of squark and gluino masses.*

## 7. Other Exotic Searches

A number of exotic extensions of the standard model have been tested with 2011 data. Here we provide just a short summary of two recent searches, and refer the interested reader to the public pages of the CMS experiment[92] for a more comprehensive list of results.

### 7.1 Search for New Resonances in the Dijet Mass Spectrum

A large class of new physics models can be tested by an inclusive search for new particles decaying to jet pairs (dijets). These models generally predict that the jets have smaller values of pseudorapidity difference than corresponding QCD backgrounds. The dijet angular ratio, defined as the ratio between the number of events with jets of similar pseudorapidity ($|\Delta\eta|<1.3$) to the number of events with jets having pseudorapidity difference $1.3<|\Delta\eta|<3.0$, is used to place limits on the cross section of these models[93].

The observed ratio is used to compute a global log-likelihood ratio of the two hypotheses (background-only and background plus signal, with varying signal mass). The $CL_s$ criterion[37-38] is used to determine 95% C.L. upper limits on the product of signal cross section, kinematical acceptance, and branching fraction to jet pairs. For the benchmark model of an excited quark, the limit is set at 3.2 TeV (see Fig. 19).

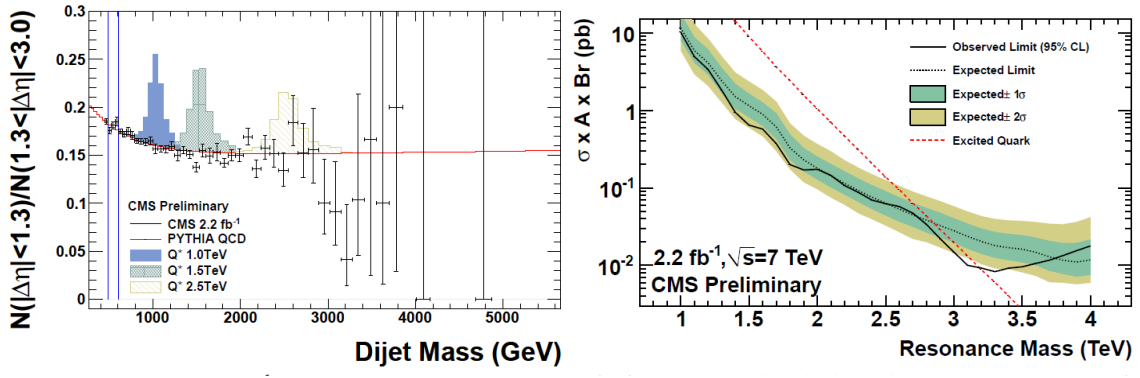

Figure 19. *Left: dijet angular ratio observed in 2.2 inverse femtobarns of CMS data (black points with error bars) compared with a QCD background shape parametrization obtained from the Pythia MC (red curve) and with the contribution expected from excited quark resonances of mass 1.0, 1.5, and 2.5 TeV. Right: upper limit on the product of production cross section, kinematic acceptance, and branching fraction to jet pairs of a dijet resonance as a function of its invariant mass. The theoretical prediction is shown as a dashed red curve.*

**7.2 Search for Microscopic Black Holes**

If the energy scale of quantum gravity phenomena were abnormally low, microscopic black holes could potentially be produced in energetic proton-proton collisions. CMS has searched for such phenomena in the context of a model with n large, flat, extra spatial dimensions[94-95], in which the scale of new physics is set by a multi-dimensional Planck scale connected with the size of the extra dimensions.

The observable quantity used to separate black hole production candidates from standard model backgrounds is the total transverse energy $S_T$ of all the detected objects in the event: jets, electrons, photons, and muons, all contributing to the sum if their transverse energy is larger than 50 GeV. The shape of the $S_T$ distribution is dominated by QCD multijet production, which is determined from a data parametrization in a control region contributed by events with only two or three objects above threshold, previously determined to be depleted of a possible signal of black hole production. The invariance properties of the variable $S_T$ on the number of objects in the final state has been thoroughly tested. $S_T$ shapes for the tested black hole production models are generated using the BLACKMAX generator[96].

Using the good agreement of observed data at high-object multiplicity and high-$S_T$ with background predictions, upper limits are set with the $CL_s$ criterion[37-38] on the cross section of black hole production in several models (see Fig. 20). The limits can be cast in the context of the considered large extra-dimensions model into lower limits in the mass of semi-classical and quantum black holes, which range between 3.8 and 5.2 TeV for a multidimensional Planck scale $M_D=4$ $TeV$ and a number of extra dimensions n ≤6, depending on a wide range of model parameters[97].

**8. Conclusions**

The CMS experiment has exploited the 5 inverse femtobarns of collisions collected in the 2011 proton-proton run of the Large Hadron Collider to produce a large number of groundbreaking results in searches for new physics and precision measurements of standard model observables. Among the most exciting of these results is certainly the narrowing down of the possible range of masses of the standard model Higgs boson, and the observation of a still insufficiently significant and yet tantalizing excess of events compatible with a 124-GeV Higgs, a mass value where the competitor ATLAS collaboration also observes a hint of Higgs decays. Data from the 8-TeV 2012 run of the LHC is expected to put the final word on the existence of that particle. Currently, however, the most striking conclusion one can draw from the set of produced results is that natural low-scale Supersymmetry is getting close to be excluded across the board of the wide SUSY parameter space; similarly, other exotics new physics models are nowhere to be seen in TeV-scale collisions.

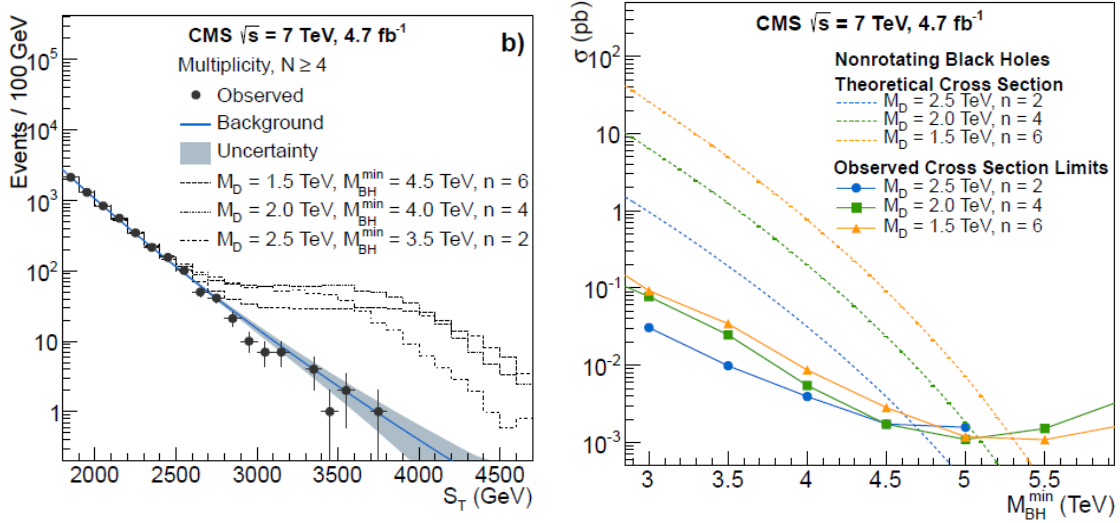

Figure 20. *Left: The distribution of the variable $S_T$ observed in CMS data with four or more final state objects (black points with error bars) compared with the background model (blue curve with error band) and the sum of background and signal contributions (empty histograms) for different values of the multi-dimensional Planck mass scale $M_D$, black hole masses, and numbers of extra dimensions n. Right: 95% C.L. upper limits on the production cross-section (in picobarns) as a function of black hole mass, for the different models shown on the left.*

## 10. Acknowledgements

We wish to congratulate our colleagues in the CERN accelerator departments for the excellent performance of the LHC machine. We thank the technical and administrative staff at CERN and other CMS institutes, and acknowledge support from: FMSR (Austria); FNRS and FWO (Belgium); Cap, CAPES, FAPERJ, and FAPESP (Brazil); MES (Bulgaria); CERN; CAS, Most, and NSFC (China); COLCIENCIAS (Colombia); MSES (Croatia); RPF (Cyprus); More, SF0690030s09 and ERDF (Estonia); Academy of Finland, MEC, and HIP (Finland); CEA and CNRS/IN2P3 (France); BMBF, DFG, and HGF (Germany); GSRT (Greece); OTKA and NKTH (Hungary); DAE and DST (India); IPM (Iran); SFI (Ireland); INFN (Italy); NRF and WCU (Korea); LAS (Lithuania); CINVESTAV, CONACYT, SEP, and UASLPFAI (Mexico); MSI (New Zealand); PAEC (Pakistan); MSHE and NSC (Poland); FCT (Portugal); JINR (Armenia, Belarus, Georgia, Ukraine, Uzbekistan); MON, Rosa tom, RAS and RFBR (Russia); MSTD (Serbia); MICINN and CPAN (Spain); Swiss Funding Agencies (Switzerland); NSC (Taipei); TUBITAK and TAEK (Turkey); STFC (United Kingdom); DOE and NSF (USA).

**References**

[1] CMS Collaboration, *Combination of SM, SM4, FP Higgs boson searches*, CMS-PAS-HIG-12-008.

[2] LHCb Collaboration, *Strong constraints on the rare decays $B_s \to \mu^+\mu^-$ and $B^0 \to \mu^+\mu^-$*, arXiv:1203.4493.

[3] CMS Collaboration, *Search for supersymmetry in pp collisions at $\sqrt{s}=7$ TeV in events with a single lepton, jets, and missing transverse momentum*, J. High Energy Phys. 08 (2011) 156, arXiv:1107.1870.

[4] CMS Collaboration, *Inclusive search for squarks and gluinos in pp collisions at $\sqrt{s}=7$ TeV*, Phys. Rev. D 85 (2012) 012004, arXiv:1107.1279.

[5] CMS Collaboration, *Search for Supersymmetry in Events with b Jets and Missing Transverse Momentum at the LHC*, J. High Energy Phys. 07 (2011) 113, arXiv:1106.3272.

[6] CMS Collaboration, *Search for New Physics with Jets and Missing Transverse Momentum in pp collisions at $\sqrt{s}=7$ TeV*, J. High Energy Phys. 08 (2011) 155, arXiv:1106.4503.

[7] CMS Collaboration, *Search for Supersymmetry in pp Collisions at $\sqrt{s}=7$ TeV in Events with Two Photons and Missing Transverse Energy*, Phys. Rev. Lett. 106 (2011) 211802, arXiv:1103.0953.

[8] CMS Collaboration, *Observation of Long-Range, Near-Side Angular Correlations in Proton-Proton Collisions at the LHC*, J. High Energy Phys. 09 (2010) 091, arXiv:1009.4122.

[9] CMS Collaboration, *Measurements of Inclusive W and Z Cross Sections in pp Collisions at $\sqrt{s}=7$ TeV*, J. High Energy Phys. 01 (2011) 080, arXiv:1012.2466.

[10] CMS Collaboration, *Upsilon production cross section in pp collisions at $\sqrt{s}=7$ TeV*, Phys. Rev. D 83 (2011) 112004, arXiv:1012.5545.

[11] CMS Collaboration, *Measurement of the $B^+$ Production Cross Section in pp Collisions at $\sqrt{s}=7$ TeV*, Phys. Rev. Lett. 106 (2011) 112001, arXiv:1101.0131.

[12] CMS Collaboration, *Measurement of the $t\bar{t}$ production cross section and the top quark mass in the dilepton channel in pp collisions at $\sqrt{s}=7$ TeV*, J. High Energy Phys. 07 (2011) 049, arXiv:1105.5661.

[13] CMS Collaboration, *The CMS experiment at the CERN LHC*, JINST 3 (2008) S08004.

[14] CMS Collaboration, *Measurements of the Inclusive W and Z Production Cross Sections in pp Collisions at $\sqrt{s}=7$ TeV with the CMS experiment*, J. High Energy Phys. 10 (2011) 132, arXiv:1107.4789.

[15] CMS Collaboration, *Jet Production Rates in Association with W and Z Bosons in pp Collisions at $\sqrt{s}=7$ TeV*, J. High Energy Phys. 01 (2012) 010, arXiv:1110.4973.

[16] CMS Collaboration, *Measurement of W$\gamma$ and Z$\gamma$ production in pp collisions at $\sqrt{s}=7$ TeV*, Phys. Lett. B 701 (2011) 535–555, arXiv:1105.2758.

[17] CMS Collaboration, *Measurement of the WW, WZ and ZZ cross sections at CMS*, CMS-PAS-EWK-11-010.

[18] J. Kubar-Andre and F. E. Paige, *Gluon corrections to the Drell–Yan model*, Phys. Rev. D19 (1979) 221.

[19] G. Altarelli, R. Ellis, and G. Martinelli, *Large Perturbative Corrections to the Drell–Yan Process in QCD*, Nucl. Phys. B 157 (1979) 461.

[20] J. Kubar et al., *QCD corrections to the Drell–Yan mechanism and the pion structure function*, Nucl. Phys. B 175 (1980) 251.

[21] P. Rijken and W. van Neerven, *Order $\alpha^2_s$ contributions to the Drell–Yan cross-section at fixed target energies*, Phys. Rev. D 51 (1995) 44.

[22] R. Hamberg, W. van Neerven, and T. Matsuura, *A complete calculation of the order $\alpha^2_s$ correction to the Drell–Yan K-factor*, Nucl. Phys. B 359 (1991) 343. Erratum-ibid. B 644 (2002) 403.

[23] W. van Neerven and and E. Zijlstra, *The $O(\alpha^2_s)$ corrected Drell–Yan K-factor in the DIS and MS scheme*, Nucl. Phys. B 382 (1992) 11. Erratum-ibid. B 680 (2004) 513.

[24] R. Harlander and W. Kilgore, *Next-to-next-to-leading order Higgs production at hadron colliders*, Phys. Rev. Lett. 88 (2002) 201801.

[25] C. Anastasiou et al., *High precision QCD at hadron colliders: Electroweak gauge boson rapidity distributions at next-to-next-to leading order*, Phys. Rev. D 69 (2004) 094008.

[26] CMS Collaboration, *Measurement of the t-channel single top quark production cross section in pp collisions at √s= 7 TeV,* Phys. Rev. Lett. 107 (2011) 091802.

[27] CMS Collaboration, *Measurement of the tt− Production Cross Section in pp Collisions at √s=7 TeV using the Kinematic Properties of Events with Leptons and Jets,* Eur. Phys. J. C 71 (2011) 1721.

[28] CMS Collaboration, *Measurement of the tt¯ production cross section and the top quark mass in the dilepton channel in pp collisions at √s =7 TeV,* J. High Energy Phys. 07 (2011) 049.

[29] M. Aliev et al., *HATHOR: HAdronic Top and Heavy quarks crOss section calculatoR*, Comput. Phys. Commun. 182 (2011) 1034.

[30] U. Langenfeld, S. Moch, and P. Uwer, *Measuring the running top-quark mass,* Phys. Rev. D 80 (2009) 054009.

[31] N. Kidonakis, *Next-to-next-to-leading soft-gluon corrections for the top quark cross section and transverse momentum distribution*, Phys. Rev. D 82 (2010) 114030.

[32] CMS Collaboration, *Measurement of the top quark mass in the muon+jets channel*, CMS-PAS-TOP-11-015

[33] E.W. N. J. A. Aguilar-Saavedra et al., *Top quark physics*, Acta Phys. Polon. B35 (2004) 2671–2694.

[34] G. Lu et al., *The Rare top quark decays t to cV in the topcolor assisted technicolor model*, Phys.Rev. D68 (2003) 015002.

[35] CMS Collaboration, *Search for Flavor Changing Neutral Currents in Top Quark Decays in pp Collisions at √s = 7 TeV,* CMS-PAS-TOP-11-028.

[36] A. J. Buras, *Minimal flavour violation and beyond: Towards a flavour code for short distance dynamics*, Acta Phys. Polon. B 41 (2010) 2487, arXiv:1012.1447.

[37] A.L. Read, *Presentation of search results: the $CL_s$ technique*, J. Phys. G: Nucl. Part. Phys. 28, 2693 (2002)

[38] T. Junk, *Confidence level computation for combining searches with small statistics*, Nucl. Instrum. Meth. A434 (1999) 435–443, arXiv:hep-ex/9902006.

[39] H1 Collaboration, *Measurement of open beauty production at HERA*, Phys. Lett. B467 (1999) 156.

[40] ZEUS Collaboration, *Measurement of charm and beauty production in deep inelastic ep scattering from decays into muons at HERA*, Eur. Phys. J. C 65 (2010) 65.

[41] CDF Collaboration, *Measurement of the bottom quark production cross-section using semileptonic decay electrons in p⁻ p collisions at √s = 1.8 TeV*, Phys. Rev. Lett. 71 (1993).

[42] D0 Collaboration, *Inclusive μ and b quark production cross-sections in p⁻ p collisions at √s = 1.8 TeV*, Phys. Rev. Lett. 74 (1995) 3548.

[43] S. Frixione et al., *Heavy quark production*, Adv. Ser. Direct. High Energy Phys. 15 (1998) 609, arXiv:hep-ph/9702287.

[44] M. Cacciari et al., *QCD analysis of first b cross-section data at 1.96-TeV*, JHEP 0407 (2004) 033, arXiv:hep-ph/0312132.

[45] M. L. Mangano, *The Saga of bottom production in p⁻p collisions*, AIP Conf. Proc. 753 (2005) 247.

[46] CMS Collaboration, *Measurement of the $B_s$ Production Cross Section with $B_s \to J/\psi\phi$ Decays in pp Collisions at √s = 7 TeV*, Phys. Rev. D 84 (2011) 052008.

[47] CMS Collaboration, *Measurement of the $B^0$ Production Cross Section in pp Collisions at √s = 7 TeV*, Phys. Rev. Lett. 106 (2011) 252001.

[48] CMS Collaboration, *Inclusive b-hadron production cross section with muons in pp collisions at √s = 7 TeV*, J. High Energy Phys. 03 (2011) 090.

[49] CMS Collaboration, *Measurement of the $B^+$ Production Cross Section in pp Collisions at √s = 7 TeV*, Phys. Rev. Lett. 106 (2011) 112001.

[50] CMS Collaboration, *Measurement of the cross section for production of b b-bar X, decaying to muons in pp collisions at √s =7 TeV*, arXiv:1203.3458.

[51] S. Frixione and B.R. Webber, *Matching NLO QCD computations and parton shower simulations*, JHEP 0206 (2002) 029, hep-ph/0204244.

[52] J. Pumplin et al., *New generation of parton distributions with uncertainties from global QCD analysis,* JHEP 0207 (2004) 012.

[53] CMS Collaboration, *J/ψ and ψ(2S) production in pp collisions at √s = 7 TeV*, J. High Energy Phys. 02 (2012) 011.

[54] F. Englert and R. Brout, *Broken symmetry and the mass of gauge vector mesons*, Phys. Rev. Lett. 13 (1964) 321–323.

[55] P. Higgs, *Broken symmetries, massless particles and gauge fields*, Phys. Lett. 12 (1964)132–133.

[56] P. Higgs, *Broken symmetries and the masses of gauge bosons*, Phys. Rev. Lett. 13 (1964) 508–509.

[57] G. Guralnik, C. Hagen, and T. Kibble, *Global conservation laws and massless particles*, Phys. Rev. Lett. 13 (1964) 585–587.

[58] P. Higgs, *Spontaneous symmetry breakdown without massless bosons*, Phys. Rev. 145 (1966) 1156–1163.

[59] T. Kibble, *Symmetry breaking in non-Abelian gauge theories*, Phys. Rev. 155 (1967) 1554–1561.

[60] CMS Collaboration, *Search for the standard model Higgs boson decaying into two photons in pp collisions at √s =7 TeV*, Phys. Lett. B 710 (2012) 403-425.

[61] CMS Collaboration, *Search for the Standard Model Higgs Boson decaying to Bottom Quarks in pp Collisions at √s =7 TeV*, Phys. Lett. B 710 (2012) 284-306.

[62] CMS Collaboration, *Search for Neutral Higgs Bosons Decaying to Tau Pairs in pp Collisions at √s = 7 TeV*, CMS PAS HIG-11-029 (2011), arXiv:1202.4083.

[63] CMS Collaboration, Search for the Higgs Boson Decaying to W+W- in the Fully Leptonic Final State in pp collisions at √s= 7 TeV, Phys. Lett. B 710 (2012) 91-113.

[64] CMS Collaboration, *Search for the standard model Higgs boson in the decay channel H→ZZ→4l in pp collisions at √s = 7 TeV*, Phys. Rev. Lett. 108 (2012) 111804.

[65] CMS Collaboration, *Search for the standard model Higgs boson in the H→ZZ→l+l−τ+τ− decay channel in pp collisions at √s =7 TeV*, CMS PAS HIG-11-028 (2011), arXiv:1202.3617.

[66] CMS Collaboration, *Search for the Higgs boson in the H → ZZ → 2l2ν channel in pp collisions at √s = 7 TeV*, CMS PAS HIG-11-026 (2011), arXiv:1202.3478.

[67] CMS Collaboration, *Search for a Higgs boson in the decay channel H→ZZ(∗)→qq̄l−l+ in pp collisions at √s = 7 TeV*, CMS PAS HIG-11-027 (2011), arXiv:1202.1416.

[68] CMS Collaboration, *Combined results of searches for the standard model Higgs boson in pp collisions at √s = 7 TeV*, Phys. Lett. B 710 (2012) 26-48.

[69] R. Barate and others (LEP Working Group for Higgs boson searches and ALEPH, DELPHI, L3, and OPAL Collaborations), *Search for the standard model Higgs boson at LEP*, Phys. Lett. B565 (2003) 61–75.

[70] ATLAS Collaboration, CMS Collaboration, and LHC Higgs Combination Group, *Procedure for the LHC Higgs boson search combination in summer 2011*, ATL-PHYS-PUB-2011-818, CMS NOTE-2011/005.

[71] E. Gross and O. Vitells, *Trial factors for the look elsewhere effect in high energy physics*, Eur. Phys. Journ. C 70 (2010) 525–530.

[72] S. Alioli et al., *NLO Higgs boson production via gluon fusion matched with shower in POWHEG*, JHEP 04 (2009) 002.

[73] P. Nason and C. Oleari, *NLO Higgs boson production via vector-boson fusion matched with shower in POWHEG*, JHEP 02 (2010) 037.

[74] T. Sjostrand, S. Mrenna, and P. Z. Skands, *PYTHIA 6.4 Physics and Manual*, JHEP 0605 (2006) 026.

[75] CMS Collaboration, *A search using multivariate techniques for a standard model Higgs boson decaying into two photons,* CMS-PAS-HIG-12-001.

[76] CMS Collaboration, *Combined results of searches for the standard model Higgs boson in pp collisions at √s = 7 TeV*, Phys. Lett. B 710 (2012) 26-48.

[77] J.Wess and B. Zumino, *Supergauge transformations in four dimensions*, Nucl. Phys. B70 (1974) 39.

[78] H. P. Nilles, *Supersymmetry, Supergravity and Particle Physics*, Phys. Reports 110 (1984) 1.

[79] H. E. Haber and G. L. Kane, *The Search for Supersymmetry: Probing Physics Beyond the Standard Model*, Phys. Reports 117 (1987) 75.

[80] R. Barbieri, S. Ferrara, and C. A. Savoy, *Gauge Models with Spontaneously Broken Local Supersymmetry*, Phys. Lett. B 119 (1982) 343.

[81] CMS Collaboration, *Search for Supersymmetry at the LHC in Events with Jets and Missing Transverse Energy,* Phys. Rev. Lett. 107 (2011) 221804.

[82] CMS Collaboration, Search *for Supersymmetry in Events with Photons and Missing Energy*, CMS-PAS-SUS-12-001.

[83] CMS Collaboration, *Search for new physics in events with same-sign dileptons, b-tagged jets and missing energy*, CMS-PAS-SUS-11-020.

[84] CMS Collaboration **,** *Search for supersymmetry in events with opposite-sign dileptons and missing energy using Artifical Neural Networks*, CMS-PAS-SUS-11-018.

[85] CMS Collaboration, *Search for supersymmetry with the razor variables at CMS*, CMS-PAS-SUS-12-005.

[86] CMS Collaboration, *Search for supersymmetry in all-hadronic events with tau leptons,* CMS-PAS-SUS-11-007.

[87] CMS Collaboration, *Search for new physics in events with b-quark jets and missing transverse energy in proton-proton collisions at 7 TeV*, CMS-PAS-SUS-11-006.

[88] CMS Collaboration, *Search for supersymmetry in all-hadronic events with MT2*, CMS-PAS-SUS-11-005.

[89] CMS Collaboration, *Search for supersymmetry in all-hadronic events with alpha_T*, CMS-PAS-SUS-11-003.

[90] CMS Collaboration**,** *Search for new physics with same-sign isolated dilepton events with jets and missing energy,* CMS-PAS-SUS-11-010.

[91] CMS Collaboration, *Multileptonic SUSY searches***,** CMS-PAS-SUS-11-013.

[92] https://twiki.cern.ch/twiki/bin/view/CMSPublic/PhysicsResultsEXO.

[93] CMS Collaboration, *Search for New Physics with the Dijet Angular Ratio*, CMS-PAS-EXO-11-026.

[94] N. Arkani-Hamed, S. Dimopoulos, and G. Dvali, *The hierarchy problem and new dimensions at a millimeter*, Phys. Lett. B 429 (1998) 263–267, arXiv:hep-ph/9803315.

[95] N. Arkani-Hamed, S. Dimopoulos, and G. Dvali, *Phenomenology, astrophysics and cosmology of theories with submillimeter dimensions and TeV scale quantum gravity*, Phys. Rev. D 59 (1999) 086004.

[96] D.-C. Dai et al., *BLACKMAX: A black-hole event generator with rotation, recoil, split branes, and brane tension*, Phys. Rev. D 77 (2008) 076007.

[97] CMS Collaboration, *Search for microscopic black holes in pp collisions at $\sqrt{s}$ =7 TeV, arXiv:1202.6396.*